\documentclass[aps,twocolumn,amsmath,amssymb]{revtex4-2}
\usepackage[dvipdfmx]{graphicx} 
\usepackage{dcolumn}
\usepackage{bm}
\usepackage{braket}
\usepackage{multirow}
\usepackage{hyperref}
\hypersetup{
colorlinks = true, 
urlcolor = blue, 
linkcolor = blue, 
citecolor = blue
}

\begin{document}

\title{
$t$-$J$ model for strongly correlated two-orbital systems: \\
Application to bilayer nickelate superconductors
}

\author{
Tatsuya Kaneko, 
Masataka Kakoi,
and Kazuhiko Kuroki
}

\affiliation{
Department of Physics, The University of Osaka, Toyonaka, Osaka 560-0043, Japan
}

\date{\today}

\begin{abstract}
We derive a $t$-$J$ model applicable to strongly correlated two-orbital systems including bilayer nickelate superconductors. Using the Schrieffer-Wolff transformation, we exclude the doubly occupied states, which raise the onsite Coulomb energy, and derive the resulting spin interactions from the two-orbital Hubbard model. We also introduce effective interactions attributed to the interorbital Coulomb interaction. To adapt the effective model to bilayer nickelates that exhibit high-temperature superconductivity, we quantitatively evaluate the strengths of the spin interactions based on the hopping parameters in La$_3$Ni$_2$O$_7$. Considering the evaluated effective interactions, we propose a simplified $t$-$J$ model for bilayer nickelate superconductors. 
\end{abstract}

\maketitle


\section{Introduction}

Understanding the electronic properties of multiorbital systems is a fundamental issue in condensed matter physics~\cite{YTokura2000,SMaekawa2004,DKhomskiia2014}. 
In correlated electron systems residing on multi-$d$ orbitals, the interplay between charge, spin, and orbital degrees of freedom gives rise to intriguing phenomena, such as unconventional superconductivity (SC)~\cite{HHosono2015,YMaeno2024} and colossal magnetoresistance~\cite{ARamirez1997,EDagotto2005}.
In contrast to single-orbital systems, we must consider the orbital degree of freedom, which brings numerical difficulties to configuring precise many-body states in strongly correlated multiorbital systems. 
Therefore, deriving effective models that project out low-probability states from the local Hilbert space is crucial in capturing the essential properties of strongly correlated multiorbital systems.  

Among intriguing phenomena in multiorbital systems, the recent discovery of high-temperature SC in bilayer nickelates has attracted significant attention~\cite{HSun2023,MWang2024}. 
La$_3$Ni$_2$O$_7$ and La$_2$PrNi$_2$O$_7$, which contain NiO$_2$ bilayer structures, show SC transitions around 80~K under pressure~\cite{JHou2023,MZhang2024,GWang2024,YZhang2024_NP,NWang2024,PPuphal2024,YUeki2025,YZhou2025,BChen2025,JLi2025}. 
SC has also been discovered in trilayer nickelates such as La$_4$Ni$_3$O$_{10}$ under pressure~\cite{HSakakibara2024_4310,QLi2024,JLi2024,YZhu2024,HNagata2024,XHuang2024,MZhang2025,EZhang2025}.  
These bulk multilayer nickelates host density wave states at ambient pressure, whereas identifying these spin and charge structures is a topic under debate~\cite{MKakoi2024_NMR,SBotzel2024,JYang2024,KChen2024,ZLiu2024,XWTi2024,TXie2024,XChen2024,LFLin2024,YWang2024,HLaBollita2024_PRM,YMeng2024,YLi2025,XRen2025,SXu2025,RKhasanov2025,DZhao2025,JLuo2025,MYashima2025,IPlokhikh_arXiv}.  
Recently, the signatures of ambient pressure SC have been reported in bilayer nickelate thin films~\cite{EKKo2025,GZhou2025}.  
These discoveries stimulate a great deal of theoretical work to elucidate the mechanism of SC in multilayer nickelates~\cite{ZLuo2023,QGYang2023,YShen2023,YFYang2023,VChristiansson2023,YZhang2023,FLechermann2023,YBLiu2023,ZLiao2023,TKaneko2024,YCao2024,HSakakibara2024_327,YZhang2024_NC,GHeier2024,RJiang2024,WWu2024,LRhodes2024,JXWang2024,BGeisler2024,HLaBollita2024_PRB,QGYang2024,CQChen2024,SRyee2024,YZhang2024_PRL,YYZheng2025,JYYou2025,MOchi2025,KYJiang2025,SKamiyama_arXiv}. 
The characteristic electronic structure in the bilayer nickelate La$_3$Ni$_2$O$_7$ is described by the correlated two-orbital model, where the $d_{3z^2-r^2}$ orbitals forming the strong interlayer bonds and the $d_{x^2-y^2}$ orbitals forming the two-dimensional conduction network are hybridized via interorbital hopping~\cite{HSakakibara2024_327,MNakata2017}.  
It is widely accepted that the interplay between multilayer geometry, multiorbitality, and strong many-body correlation gives rise to the complicated electronic structures in multilayer nickelate superconductors. 

In the studies of unconventional SC in strongly correlated systems, the $t$-$J$ model, in which the doubly occupied ($\uparrow\downarrow$) state is excluded, has often been used to grasp the essential characteristics of mobile carriers on magnetic backgrounds.   
The single-orbital $t$-$J$ model has been used to explain the superconducting properties, such as $d$-wave SC and phase diagram, of cuprate superconductors~\cite{PLee2006,MOgata2008}. 
In terms of numerical studies, the computation of the $t$-$J$ model is less expensive compared with the Hubbard model because the number of local electron configurations of the $t$-$J$ model ($\circ$, $\uparrow$, $\downarrow$) is less than that of the Hubbard model ($\circ$, $\uparrow$, $\downarrow$, $\uparrow\downarrow$). 
Hence, the $t$-$J$ model enables us to compute wave functions of large-size systems inaccessible by the Hubbard model and can enlarge the applicability of many-body solvers, such as the exact diagonalization and density-matrix renormalization group methods.   
The mechanism of SC in bilayer nickelates has been studied using various types of $t$-$J$ models~\cite{HLange2024,XZQu2024,CLu2024_1,ZPan2024,JChen2024,JXZhang2024,HSchlomer2024,MBejas2025,YTian2025,PBorchia_arXiv,CLu_arXiv,MKakoi2024_DMRG,ZLuo2024,CLu2024_2,XZQu_arXiv,ZShao_arXiv,GDuan_arXiv,HOh2023,JRXue2024,HYang2024,HOh2025,HYang2025}. 
Although the SC properties have been studied in two-orbital $t$-$J$ models, the valid spin interactions due to multiorbitality are not fully considered in the simplified $t$-$J$ models in the previous studies. 
Therefore, a comprehensive derivation of an effective model from the two-orbital Hubbard model is desired.   

In this paper, we derive an $t$-$J$ model applicable to strongly correlated two-orbital systems including bilayer nickelate superconductors.  
First, we classify the hopping processes in the two-orbital Hubbard model with Hund's coupling to evaluate the energy changes associated with single-particle hopping. 
Then, using the Schrieffer-Wolff transformation, we derive effective spin interactions as a consequence of excluding the doubly occupied states that raise the onsite Coulomb energy.   
Furthermore, we introduce effective interactions, equivalent to the Kugel-Khomskii--type interactions, originating from the interorbital Coulomb interaction. 
To adapt our effective model to bilayer nickelate superconductors, we present the spin interactions based on the hopping parameters evaluated in La$_3$Ni$_2$O$_7$. 
Finally, considering the evaluated effective interactions, we propose a simplified $t$-$J$ model for bilayer nickelate superconductors. 

This paper is organized as follows. 
In Sec.~\ref{sec:TOHM}, we introduce the two-orbital Hubbard model and its eigenstates at a single site. 
In Sec.~\ref{sec:model_I}, we derive effective interactions from the second-order perturbation theory when doubly occupied orbitals are excluded. 
In Sec.~\ref{sec:model_II}, we formulate effective interactions obtained by excluding the empty state, which is unfavorable when the occupancy of two orbitals is larger than one.  
In Sec.~\ref{sec:model_LNO}, we evaluate the strength of the effective interactions based on the hopping parameters for La$_3$Ni$_2$O$_7$ and suggest a simplified $t$-$J$ model for bilayer nickelate superconductors. 
Section~\ref{sec:summary} summarizes our study.


\section{Two-orbital Hubbard model} \label{sec:TOHM}

\subsection{Hamiltonian}

We consider an effective model derived from the two-orbital Hubbard model. 
The Hamiltonian of the two-orbital Hubbard model reads
\begin{align}
\hat{\mathcal{H}} = \hat{\mathcal{H}}_0 + \hat{\mathcal{T}}
\end{align}
with the onsite term including the local Coulomb interactions 
\begin{align}
&\hat{\mathcal{H}}_0 
= \Delta \sum_{\bm{R}} \hat{n}_{\bm{R},b} 
+ U \sum_{\bm{R}} \sum_{\mu} \hat{n}_{\bm{R},\mu,\uparrow}  \hat{n}_{\bm{R},\mu,\downarrow} 
\notag \\
&+ U' \sum_{\bm{R}} \hat{n}_{\bm{R},a} \hat{n}_{\bm{R},b} 
-2 J_{\rm H} \sum_{\bm{R}} \left( \hat{\bm{s}}_{\bm{R},a} \cdot \hat{\bm{s}}_{\bm{R},b} + \frac{1}{4} \hat{n}_{\bm{R},a} \hat{n}_{\bm{R},b} \right)  
\label{eq:H_TOHM}
\end{align}
and the hopping term
\begin{align}
\hat{\mathcal{T}} = -\sum_{\bm{R},\bm{R}'} \sum_{\mu,\nu} \sum_{\sigma} t^{\mu\nu}_{\bm{R}\bm{R}'} \hat{c}^{\dag}_{\bm{R},\mu,\sigma} \hat{c}_{\bm{R}',\nu,\sigma}. 
\end{align}
$\hat{c}^{\dag}_{\bm{R},\mu,\sigma}$ ($\hat{c}_{\bm{R},\mu,\sigma}$) is the creation (annihilation) operator of an electron with spin $\sigma$~$(=\uparrow,\downarrow)$ in orbital $\mu$~$(=a,b)$ at position $\bm{R}$. 
$\hat{n}_{\bm{R},\mu} = \sum_{\sigma} \hat{n}_{\bm{R},\mu,\sigma} = \sum_{\sigma} \hat{c}^{\dag}_{\bm{R},\mu,\sigma} \hat{c}_{\bm{R},\mu,\sigma}$ is the number operator, and  $\hat{\bm{s}}_{\bm{R},\mu} = (1/2) \sum_{\sigma,\sigma'} \hat{c}^{\dag}_{\bm{R},\mu,\sigma}  \bm{\sigma}_{\sigma\sigma'} \hat{c}_{\bm{R},\mu,\sigma'}$ is the spin operator, where $\bm{\sigma}=(\sigma^x,\sigma^y, \sigma^z)$ is the vector of Pauli matrices. 
$t^{\mu\nu}_{\bm{R}\bm{R}'}$ is the hopping integral between $(\bm{R},\mu)$ and $(\bm{R}',\nu)$. 
Here, we assume $|t^{ab}_{\bm{R}\bm{R}'}| = |t^{ba}_{\bm{R}\bm{R}'}|$. 
$\Delta$ is the energy level difference between two orbitals. 
For example, the $a$ and $b$ orbitals can correspond to the $d_{3z^2-r^2}$ and $d_{x^2-y^2}$ orbitals in the bilayer nickelate superconductors, where the energy level of the $a = d_{3z^2-r^2}$ orbital is lower than that of the $b = d_{x^2-y^2}$ orbital due to the crystal-field splitting~\cite{ZLuo2023,HSakakibara2024_327}. 
$U$ and $U'$ are the intraorbital and interorbital Coulomb interactions, respectively. 
$J_{\rm H}$~$(>0)$ is Hund's coupling (i.e., ferromagnetic interorbital spin interaction).   
We assume that the electron filling is less than half (i.e., $\braket{\hat{n}_{\bm{R},a}} + \braket{\hat{n}_{\bm{R},b}} \le 2$). 

Here, we derive an effective model of the two-orbital Hubbard model without the pair-hopping term because the pair-hopping term does not change the single-site eigenstates used in the ground-state configuration of the effective model with no doubly occupied orbital. 
The roles of the pair-hopping term are summarized in Appendix~\ref{appendix_A}. 
Although the formulas become complicated, the derivation of an effective model incorporating pair hopping can be conducted using essentially the same procedure shown later.

\subsection{Eigenstates of the onsite term $\hat{\mathcal{H}}_0$}

\renewcommand{\arraystretch}{1.8}
\begin{table}[!b]
\caption{Eigenstates and eigenenergies of $\hat{\mathcal{H}}_0$.}
\centering
\begin{tabular}{lcc}
\hline \hline
Symbol & State & Energy 
\\ 
\hline 
$0$ & $\ket{0}$ & $0$ 
\\ 
$\sigma_a$ & $\hat{c}^{\dag}_{a,\sigma} \ket{0}$ & $0$ 
\\ 
$\sigma_b$ & $\hat{c}^{\dag}_{b,\sigma} \ket{0}$ & $\Delta$ 
\\ 
${\rm S}$ & $\frac{1}{\sqrt{2}} \bigl( \hat{c}^{\dag}_{a,\uparrow}\hat{c}^{\dag}_{b,\downarrow} \!-\! \hat{c}^{\dag}_{a,\downarrow}\hat{c}^{\dag}_{b,\uparrow} \bigr) \ket{0}$ & $U' + J_{\rm H} + \Delta$ 
\\ 
${\rm T}_{+}$ & $\hat{c}^{\dag}_{a,\uparrow}\hat{c}^{\dag}_{b,\uparrow} \ket{0}$ &  $U' - J_{\rm H} + \Delta$ 
\\ 
${\rm T}_0$ & $\frac{1}{\sqrt{2}} \bigl( \hat{c}^{\dag}_{a, \uparrow}\hat{c}^{\dag}_{b,\downarrow} \!+\! \hat{c}^{\dag}_{a,\downarrow}\hat{c}^{\dag}_{b,\uparrow} \bigr) \ket{0}$ & $U' - J_{\rm H} + \Delta$ 
\\ 
${\rm T}_{-}$ & $\hat{c}^{\dag}_{a,\downarrow}\hat{c}^{\dag}_{b,\downarrow} \ket{0}$ & $U' - J_{\rm H} + \Delta$ 
\\ 
${\rm D}_a$ & $\hat{c}^{\dag}_{a,\uparrow}\hat{c}^{\dag}_{a,\downarrow} \ket{0}$ & $U$ 
\\ 
${\rm D}_b$ & $\hat{c}^{\dag}_{b,\uparrow}\hat{c}^{\dag}_{b,\downarrow} \ket{0}$ & $U+2\Delta$ 
\\
$\sigma_b{\rm D}_a$ & $\hat{c}^{\dag}_{b,\sigma} \hat{c}^{\dag}_{a,\uparrow}\hat{c}^{\dag}_{a,\downarrow} \ket{0}$ & $U + 2U' - J_{\rm H} + \Delta$ 
\\ 
$\sigma_a{\rm D}_b$ & $\hat{c}^{\dag}_{a,\sigma} \hat{c}^{\dag}_{b,\uparrow}\hat{c}^{\dag}_{b,\downarrow} \ket{0}$ & $U + 2U' - J_{\rm H} + 2\Delta$ 
\\ 
${\rm D}_a{\rm D}_b$ & $\hat{c}^{\dag}_{a,\uparrow}\hat{c}^{\dag}_{a,\downarrow}  \hat{c}^{\dag}_{b,\uparrow}\hat{c}^{\dag}_{b,\downarrow} \ket{0}$ & $2U + 4U' - 2 J_{\rm H} + 2 \Delta$ 
\\ 
\hline \hline
\end{tabular}
\label{table1}
\end{table}
\renewcommand{\arraystretch}{1}

We derive an effective model assuming that the energy scale of $\hat{\mathcal{H}}_0$ is larger than the hopping parameter $t^{\mu\nu}_{\bm{R}\bm{R}'}$. 
Here, we summarize the eigenstates of $\hat{\mathcal{H}}_0$, which is the unperturbed term in the strong-coupling expansion. 
This section considers the eigenstates at a single site, and we do not explicitly indicate position $\bm{R}$. 
There are 16 (=~4$^2$) states at a single site with two orbitals.  Table~\ref{table1} presents all eigenstates and eigenenergies of $\hat{\mathcal{H}}_0$. 
Since $\hat{\mathcal{H}}_0$ includes the spin-flip term in Hund's coupling, the eigenstates of the $S^z=0$ states with two singly occupied orbitals are given by the spin-singlet state $\ket{{\rm S}} = \frac{1}{\sqrt{2}} ( \hat{c}^{\dag}_{a,\uparrow}\hat{c}^{\dag}_{b,\downarrow} - \hat{c}^{\dag}_{a,\downarrow}\hat{c}^{\dag}_{b,\uparrow} ) \ket{0}$ and the spin-triplet state $\ket{{\rm T}_0} = \frac{1}{\sqrt{2}} ( \hat{c}^{\dag}_{a,\uparrow}\hat{c}^{\dag}_{b,\downarrow} + \hat{c}^{\dag}_{a,\downarrow}\hat{c}^{\dag}_{b,\uparrow} ) \ket{0}$. 
 
We introduce projection operators to identify each eigenstate.   
For example, $\ket{\uparrow_{a}} = \hat{c}^{\dag}_{a,\uparrow} \ket{0}$ can be identified by the projection operator $\hat{P}(\uparrow_{a}) = \hat{n}_{a,\uparrow} \left( 1 - \hat{n}_{a,\downarrow} \right) \left( 1 - \hat{n}_{b,\uparrow} \right) \left( 1 - \hat{n}_{b,\downarrow} \right)$.  
The other projection operators can be defined similarly. 
Note that the projection operators of $\ket{{\rm S}}$ and $\ket{{\rm T}_0}$ include $\hat{\bm{s}}_{a} \cdot \hat{\bm{s}}_{b}$.  
Specifically, the projection operator of the spin-singlet state is   
\begin{align}
\hat{P}({\rm S}) & = 
\frac{1}{4} \left( \hat{n}_{a,\uparrow} \!+\! \hat{n}_{a,\downarrow} \!-\! 2 \hat{n}_{a,\uparrow} \hat{n}_{a,\downarrow} \right) \left( \hat{n}_{b,\uparrow} \!+\! \hat{n}_{b,\downarrow} \!-\! 2 \hat{n}_{b,\uparrow} \hat{n}_{b,\downarrow} \right) 
\notag \\
&- \hat{\bm{s}}_{a} \cdot \hat{\bm{s}}_{b}, 
\end{align}
while the projection operator of the spin-triplet states is
\begin{align}
\hat{P}({\rm T}) & = 
\frac{3}{4} \left( \hat{n}_{a,\uparrow} \!+\! \hat{n}_{a,\downarrow} \!-\! 2 \hat{n}_{a,\uparrow} \hat{n}_{a,\downarrow} \right) \left( \hat{n}_{b,\uparrow} \!+\! \hat{n}_{b,\downarrow} \!-\! 2 \hat{n}_{b,\uparrow} \hat{n}_{b,\downarrow} \right) 
\notag \\
&+ \hat{\bm{s}}_{a} \cdot \hat{\bm{s}}_{b}. 
\end{align}
Note that $\hat{P}({\rm T}) = \hat{P}({\rm T}_{+}) + \hat{P}({\rm T}_0) + \hat{P}({\rm T}_{-})$.  
If one of the spin-triplet states needs to be identified, we must use the projection operator for each state. 
For example, $\hat{P}({\rm T}_{+}) = \hat{n}_{a,\uparrow} \left( 1 - \hat{n}_{a,\downarrow} \right) \hat{n}_{b,\uparrow} \left( 1 - \hat{n}_{b,\downarrow} \right)$ for $\ket{{\rm T}_{+}}$. 
For later convenience, we introduce the projector operator $\hat{P}(\mu) = \hat{P}(\uparrow_{\mu}) + \hat{P}(\downarrow_{\mu})$, which identifies the singly occupied state on orbital $\mu$ regardless of spin $\sigma$. 
This operator $\hat{P}(\mu)$ can be written as 
\begin{align}
\hat{P}(\mu) = \left( \hat{n}_{\mu,\uparrow} + \hat{n}_{\mu,\downarrow} - 2 \hat{n}_{\mu,\uparrow} \hat{n}_{\mu,\downarrow} \right) \left( 1 - \hat{n}_{\bar{\mu},\uparrow} \right) \left( 1 - \hat{n}_{\bar{\mu},\downarrow} \right), 
\end{align}
where $\bar{\mu}$ indicates the opposite orbital of $\mu$ (i.e., $\bar{a}=b$ and $\bar{b}=a$).


\section{Exclusion of doubly occupied orbitals} \label{sec:model_I}

When the onsite Coulomb interaction $U$ is sufficiently larger than the hopping parameter $t^{\mu\nu}_{\bm{R}\bm{R}'}$, the doubly occupied state, in which two electrons occupy a single orbital, is unlikely to be created. 
In this situation, we can approximately omit low-probability states with doubly occupied orbitals from the ground-state configuration.  
We include their contributions as effective spin interactions caused by second-order perturbations.

\subsection{Schrieffer-Wolff transformation}

Assuming $U\gg t^{\mu\nu}_{\bm{R}\bm{R}'}$, we derive an effective model as a consequence of excluding doubly occupied orbitals.  
To exclude the doubly occupied orbital with energy of $U$, we use the transformation~\cite{AMacDonald1988}
\begin{align}
\hat{\mathcal{H}}' = e^{i\hat{\mathcal{S}}} \hat{\mathcal{H}} e^{-i\hat{\mathcal{S}}}
= \hat{\mathcal{H}} + \left[ i\hat{\mathcal{S}} , \hat{\mathcal{H}} \right] + \frac{1}{2} \left[ i\hat{\mathcal{S}} , \left[ i\hat{\mathcal{S}} ,  \hat{\mathcal{H}} \right] \right] + \cdots. 
\label{eq:SWT_H'}
\end{align}
We describe an expansion of $\hat{\mathcal{S}}$ in powers of $t^{\mu\nu}_{\bm{R}\bm{R}'}$ as $\hat{\mathcal{S}} = \hat{\mathcal{S}}^{(1)} + \hat{\mathcal{S}}^{(2)} + \cdots$. 
When $U\gg t^{\mu\nu}_{\bm{R}\bm{R}'}$, higher-order contributions on the order of $(t^{\mu\nu}_{\bm{R}\bm{R}'}/U)^n$ become negligible. 
To derive an effective model without doubly occupied orbitals, we decompose the hopping term into
\begin{align}
\hat{\mathcal{T}} = \hat{\mathcal{T}}_{0} + \hat{\mathcal{T}}_{+1} + \hat{\mathcal{T}}_{-1}  , 
\end{align}
where the hopping term $\hat{\mathcal{T}}_0$ does not change the number of doubly occupied orbitals $N_d$ [see Fig.~\ref{fig1}(a)], while the hopping term $\hat{\mathcal{T}}_{\pm 1}$ does change $N_d$, i.e., $N_d \rightarrow N_d \pm 1$ [see Fig.~\ref{fig1}(b)]. 
Here, we derive an effective model at $N_d=0$. 
In order to get rid of $\hat{\mathcal{T}}_{+1} + \hat{\mathcal{T}}_{-1}$ that changes $N_d$, we set $\hat{\mathcal{S}}^{(1)}$ to satisfy 
\begin{align}
\hat{\mathcal{T}}_{+1} + \hat{\mathcal{T}}_{-1} + \left[ i\hat{\mathcal{S}}^{(1)} , \hat{\mathcal{H}}_0 \right] = 0. 
\label{eq:SWT_S1}
\end{align}
This setup leads to the Hamiltonian 
\begin{align}
\hat{\mathcal{H}}' 
= \hat{\mathcal{H}}_0 + \hat{\mathcal{T}}_0 
+ \frac{1}{2} \left[ i\hat{\mathcal{S}}^{(1)} , \left( \hat{\mathcal{T}}_{+1} + \hat{\mathcal{T}}_{-1} \right) \right]  
+ \cdots.
\end{align} 
In this paper, we focus on the interactions derived from the second-order term $(1/2)[ i\hat{\mathcal{S}}^{(1)}, ( \hat{\mathcal{T}}_{+1} + \hat{\mathcal{T}}_{-1}  ) ]$.
 
Equation~(\ref{eq:SWT_S1}) can be satisfied by 
\begin{align}
\left[ \hat{\mathcal{H}}_0 , i\hat{\mathcal{S}}^{(1)}_{\pm 1} \right] = \hat{\mathcal{T}}_{\pm 1} 
\label{eq:SWT_S1_pm}
\end{align}
with 
\begin{align}
i\hat{\mathcal{S}}^{(1)} = i \hat{\mathcal{S}}^{(1)}_{+1} + i\hat{\mathcal{S}}^{(1)}_{-1}. 
\end{align}
An effective model in the second-order perturbation theory can be obtained by $i\hat{\mathcal{S}}^{(1)}_{\pm 1}$ satisfying Eq.~(\ref{eq:SWT_S1_pm}). 
Equation (\ref{eq:SWT_S1_pm}) implies that the operator $i\hat{\mathcal{S}}^{(1)}_{\pm 1}$ changes the doublon number from $N_d$ to $N_d\pm 1$. 
The Hamiltonian in Eq.~(\ref{eq:SWT_H'}) contains $\bigl[ i\hat{\mathcal{S}}^{(1)}_{\pm}, \hat{\mathcal{T}}_0 \bigr]$ that changes $N_d$ as a second-order term, whereas this term can be eliminated by an optimal choice of $i\hat{\mathcal{S}}^{(2)}$~\cite{AMacDonald1988}.

\subsection{Second-order perturbation theory}

\begin{figure}[b]
\begin{center}
\includegraphics[width=\columnwidth]{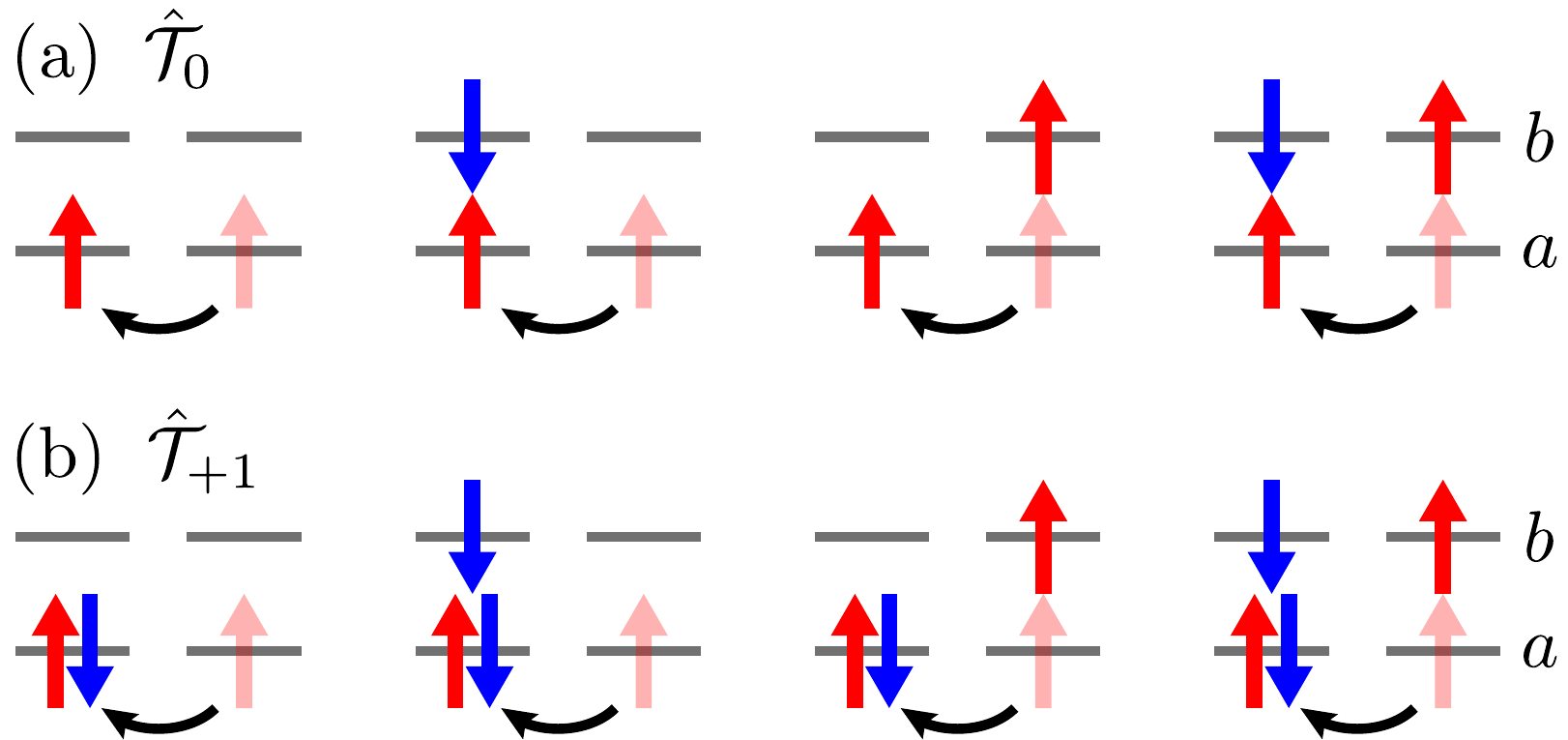} 
\caption{Hoppings in (a) $\hat{\mathcal{T}}_0$ and (b) $\hat{\mathcal{T}}_{+1}$.} 
\label{fig1}
\end{center}
\end{figure}

We introduce the specific forms of $\hat{\mathcal{T}}_0$ and $\hat{\mathcal{T}}_{\pm 1}$ for deriving an effective model without doubly occupied orbitals.  
To restrict the single-site states, we define the projection operator 
\begin{align}
\hat{P}_{\bm{R},\mu} = \hat{P}_{\bm{R}}(\mu) + \hat{P}_{\bm{R}}({\rm S}) +  \hat{P}_{\bm{R}}({\rm T}) 
= \sum_{\xi} \hat{P}_{\bm{R}}(\xi) . 
\end{align}
This operator can project out the states that include doubly occupied orbitals (D$_{\rm \mu}$). 
Using $\hat{P}_{\bm{R},\mu}$, the hopping terms are given by 
\begin{align}
&\hat{\mathcal{T}}_0 = -\sum_{\bm{R},\bm{R}'} \sum_{\mu,\nu} \sum_{\sigma} 
t^{\mu\nu}_{\bm{R}\bm{R}'}  
\hat{P}_{\bm{R},\mu} 
\hat{c}^{\dag}_{\bm{R},\mu,\sigma} \hat{c}_{\bm{R}',\nu,\sigma}   
\hat{P}_{\bm{R}',\nu}, 
\label{eq:T0} \\
&\hat{\mathcal{T}}_{+1} = -\sum_{\bm{R},\bm{R}'} \sum_{\mu,\nu} \sum_{\sigma} t^{\mu\nu}_{\bm{R}\bm{R}'}  
\hat{c}^{\dag}_{\bm{R},\mu,\sigma} \hat{c}_{\bm{R}',\nu,\sigma}  
\hat{P}_{\bm{R},\mu} \hat{P}_{\bm{R}',\nu} , 
\label{eq:T+} \\
&\hat{\mathcal{T}}_{-1} = -\sum_{\bm{R},\bm{R}'} \sum_{\mu,\nu} \sum_{\sigma} t^{\mu\nu}_{\bm{R}\bm{R}'}  
\hat{P}_{\bm{R},\mu} \hat{P}_{\bm{R}',\nu} 
\hat{c}^{\dag}_{\bm{R},\mu,\sigma} \hat{c}_{\bm{R}',\nu,\sigma} ,
\label{eq:T-}
\end{align}
where $\hat{\mathcal{T}}_{-1} = (\hat{\mathcal{T}}_{+1})^{\dag}$. 
Several hopping patterns in $\hat{\mathcal{T}}_0$ and $\hat{\mathcal{T}}_{+1}$ are schematically shown in Fig.~\ref{fig1}. 
$\hat{\mathcal{T}}_0$ does not change the number of doubly occupied orbitals [see Fig.~\ref{fig1}(a)]. 
Since $\hat{P}_{\bm{R},\mu} \hat{P}_{\bm{R}',\nu}$ in $\hat{\mathcal{T}}_{+1}$ indicates that the $\mu$ orbital at $\bm{R}$ is singly occupied, the hopping operator $\hat{c}^{\dag}_{\bm{R},\mu,\sigma} \hat{c}_{\bm{R}',\nu,\sigma}$ in $\hat{\mathcal{T}}_{+1}$ creates one doubly occupied orbital (D$_{\mu}$) at $\bm{R}$ [see Fig.~\ref{fig1}(b)]. 

\renewcommand{\arraystretch}{1.8}
\begin{table}[t]
\caption{Energy change due to single-particle hopping, where $\Delta_{\mu\nu} = \Delta \left( \delta_{\mu,b} - \delta_{\nu,b} \right)$.}
\centering
\begin{tabular}{c}
\begin{tabular}{ccc}
\hline \hline
$\xi$ ($\bm{R}$) & $ \xi'$ ($\bm{R}'$) & $\Delta E(\xi,\xi')$ 
\\ 
\hline 
$\mu$ & $\nu$ & $U + \Delta_{\mu\nu}$ 
\\ 
$\mu$ & ${\rm S}$ & $U - U' - J_{\rm H} + \Delta_{\mu\nu}$ 
\\ 
${\rm S}$ & $\nu$ & $U + U' - 2J_{\rm H}  + \Delta_{\mu\nu}$ 
\\ 
$\mu$ & ${\rm T}$ & $U - U' + J_{\rm H} + \Delta_{\mu\nu}$ 
\\ 
${\rm T}$ & $\nu$ & $U + U' + \Delta_{\mu\nu}$  
\\ 
${\rm S}$ & ${\rm S}$ & $U - 3J_{\rm H} + \Delta_{\mu\nu}$ 
\\ 
${\rm S}$ & ${\rm T}$ & $U - J_{\rm H} + \Delta_{\mu\nu} $ 
\\ 
 ${\rm T}$  & ${\rm S}$ & $U - J_{\rm H} + \Delta_{\mu\nu} $
\\ 
${\rm T}$ & ${\rm T}$ & $U + J_{\rm H} + \Delta_{\mu\nu}$
\\ 
\hline \hline
\end{tabular}
\end{tabular}
\label{table2}
\end{table}
\renewcommand{\arraystretch}{1}

With respect to the hopping term $\hat{\mathcal{T}}_{+1}$, we find that $i \hat{\mathcal{S}}^{(1)}_{+1} $ satisfying $\bigl[ \hat{\mathcal{H}}_0 , i\hat{\mathcal{S}}^{(1)}_{+1} \bigr] = \hat{\mathcal{T}}_{+1}$ in Eq.~(\ref{eq:SWT_S1_pm}) is given by 
\begin{align}
i\hat{\mathcal{S}}^{(1)}_{+1} = -\sum_{\bm{R},\bm{R}'} \sum_{\mu,\nu} \sum_{\sigma} \sum_{\xi,\xi'} 
\frac{t^{\mu\nu}_{\bm{R}\bm{R}'}}{\Delta E(\xi,\xi')} 
\hat{c}^{\dag}_{\bm{R},\mu,\sigma} \hat{c}_{\bm{R}',\nu,\sigma} &
\notag \\
\times \hat{P}_{\bm{R}}(\xi) \hat{P}_{\bm{R}'}(\xi') &. 
\label{eq:SWT_S1_+1}
\end{align}
The derivation of $i \hat{\mathcal{S}}^{(1)}_{+1}$ is given in Appendix~\ref{appendix_B}.
$\Delta E(\xi,\xi')$ is the energy change associated with the single-particle hopping between the $\xi$ and $\xi'$ initial states. 
$\Delta E(\xi, \xi')$ for all $N_d=0 \rightarrow N_d=1$ processes are presented in Table~\ref{table2}, where 
\begin{align}
\Delta_{\mu\nu} = \Delta \left( \delta_{\mu,b} - \delta_{\nu,b} \right). 
\end{align}
$\hat{\mathcal{S}}^{(1)}_{-1}$ can be obtained by $\hat{\mathcal{S}}^{(1)}_{-1}=\hat{\mathcal{S}}^{(1)\dag}_{+1}$. 
Using $i\hat{\mathcal{S}}^{(1)}_{+1}$ in Eq.~(\ref{eq:SWT_S1_+1}), the effective Hamiltonian at the second order is given by 
\begin{align}
\hat{\mathcal{H}}^{(2)}_{\rm eff} = - \frac{1}{2}\hat{\mathcal{T}}_{-1} \left( i \hat{\mathcal{S}}^{(1)}_{+1} \right) + {\rm H.c.}
\label{eq:Heff_2nd}
\end{align}
in the $N_d=0$ sector. 

This effective Hamiltonian generates many effective interactions between the single spin  $\sigma_{\mu}$ ($S=\frac{1}{2}$), spin-singlet S ($S=0$), and spin-triplet T ($S=1$) states. 
Equation~(\ref{eq:Heff_2nd}) contains two- and three-site interactions. 
The specific forms of the two- and three-site terms are presented in Eqs.~(\ref{eq:Heff_2nd_2site}) and (\ref{eq:Heff_2nd_3site}) in Appendix~\ref{appendix_B}. 
These equations enable us to obtain all effective interactions.

\subsection{Spin interactions}

We introduce several key interactions in Eq.~(\ref{eq:Heff_2nd}). 
We can approximately omit the spin-singlet (S) in the ground-state configuration when $J_{\rm H}$ is sufficiently large.  
Here, we introduce the two-site spin interactions between the spin-$\frac{1}{2}$ ($\sigma_{\mu}$) and spin-1 (T) states. 
If the effective model with all contributions should be considered, we must incorporate the interactions with the spin-singlet (S). 
We present all two-site interactions involving the spin-singlet state (S) in Appendix~\ref{appendix_C}. 

To describe the effective spin interactions, we introduce the spin-1 operator defined as 
\begin{align}
\hat{\bm{S}} = \sum_{\zeta,\zeta'} \left(\bm{S}\right)_{\zeta\zeta'} \ket{{\rm T}_{\zeta}}\bra{{\rm T}_{\zeta'}}, 
\end{align}
where $\bm{S}=(S^x,S^y,S^z)$ is the vector of the 3$\times$3 matrices
\begin{gather}
S^{x} = \frac{1}{\sqrt{2}}
\left(
\begin{array}{ccc}
0 & 1 & 0 \\
1 & 0 & 1 \\
0 & 1 & 0 
\end{array}
\right), 
\;\;\;
S^{y} = \frac{1}{\sqrt{2}}
\left(
\begin{array}{ccc}
0 &-i & 0 \\
i & 0 &-i \\
0 & i & 0 
\end{array}
\right), 
\notag \\
S^{z} = 
\left(
\begin{array}{ccc}
1 & 0 & 0 \\
0 & 0 & 0 \\
0 & 0 &-1 
\end{array}
\right). 
\end{gather}
These 3$\times$3 matrices are spanned by $\ket{{\rm T}_{+}}$, $\ket{{\rm T}_{0}}$, and $\ket{{\rm T}_{-}}$. 
The raising and lowering operators are given by $\hat{S}^{\pm}=\hat{S}^x\pm i\hat{S}^y$.

\subsubsection{spin-$\frac{1}{2}$ -- spin-$\frac{1}{2}$ interaction}

When both $\bm{R}$ and $\bm{R}'$ sites are the singly occupied states ($\mu$ and $\nu$), the $(\bm{R}',\nu) \rightarrow (\bm{R},\mu) \rightarrow (\bm{R}',\nu)$ process [Fig.~\ref{fig2}(a)] gives the spin-$\frac{1}{2}$ -- spin-$\frac{1}{2}$ interaction 
\begin{align} 
\hat{\mathcal{H}}^{(2)}_{J,\frac{1}{2}\frac{1}{2}}
= \sum_{\langle \bm{R},\bm{R}' \rangle} \sum_{\mu,\nu} 
J^{\mu\nu}_{\bm{R}\bm{R}'}
&\hat{P}_{\bm{R}}(\mu) \hat{P}_{\bm{R}'}(\nu) 
\left( \hat{\bm{s}}_{\bm{R},\mu} \cdot \hat{\bm{s}}_{\bm{R}',\nu} - \frac{1}{4} \right) 
\notag \\
&\times \hat{P}_{\bm{R}}(\mu) \hat{P}_{\bm{R}'}(\nu) 
\end{align}
with 
\begin{align}
J^{\mu\nu}_{\bm{R}\bm{R}'} = 
\frac{2|t^{\mu\nu}_{\bm{R}\bm{R}'}|^2}{U \!+\! \Delta_{\mu\nu}} 
+\frac{2|t^{\mu\nu}_{\bm{R}\bm{R}'}|^2}{U \!-\! \Delta_{\mu\nu}}  , 
\label{eq:J_munu}
\end{align}
where $\langle \bm{R},\bm{R}' \rangle$ indicates a pair of sites $\bm{R}$ and $\bm{R}'$. 
In addition, the $(\bm{R}',\nu) \rightarrow (\bm{R},\mu) \rightarrow (\bm{R}',\bar{\nu})$ process [Fig.~\ref{fig2}(a)] is also possible due to the interorbital hopping $t^{ab}_{\bm{R}\bm{R}'}$. 
This process leads to the correlated interorbital hopping term, whose Hamiltonian is given by 
\begin{align}
&\hat{\mathcal{H}}^{(2)}_{I,\frac{1}{2}\frac{1}{2}}
= \sum_{\langle \bm{R},\bm{R}' \rangle} \sum_{\mu,\nu}
I^{\mu, \bar{\nu}\nu}_{\bm{R}\bm{R}'} 
\hat{P}_{\bm{R}}(\mu) \hat{P}_{\bm{R}'}(\bar{\nu}) 
\times
\notag \\
&\left[
\sum_{\sigma,\sigma'} \hat{c}^{\dag}_{\bm{R}',\bar{\nu},\sigma'}
\left( \hat{\bm{s}}_{\bm{R},\mu} \! \cdot \!  \frac{\bm{\sigma}_{\sigma'\sigma}}{2} \! - \! \frac{1}{4} \delta_{\sigma',\sigma}\right) 
\hat{c}_{\bm{R}',\nu,\sigma} 
\right]
\hat{P}_{\bm{R}}(\mu) \hat{P}_{\bm{R}'}(\nu) 
\notag \\
& \qquad \qquad + \left( \bm{R} \leftrightarrow \bm{R}' \right) 
\end{align}
with 
\begin{align}
& I^{\mu,\bar{\nu}\nu}_{\bm{R}\bm{R}'}  
= \frac{t^{\bar{\nu}\mu}_{\bm{R}'\bm{R}} t^{\mu\nu}_{\bm{R}\bm{R}'}}{U \!+\! \Delta_{\mu\nu}} 
+ \frac{t^{\bar{\nu}\mu}_{\bm{R}'\bm{R}} t^{\mu\nu}_{\bm{R}\bm{R}'}}{U \!+\! \Delta_{\mu\bar{\nu}}} . 
\label{eq:I_mununu}
\end{align}

\subsubsection{spin-$\frac{1}{2}$ -- spin-$1$ interaction}

\begin{figure}[t]
\begin{center}
\includegraphics[width=0.95\columnwidth]{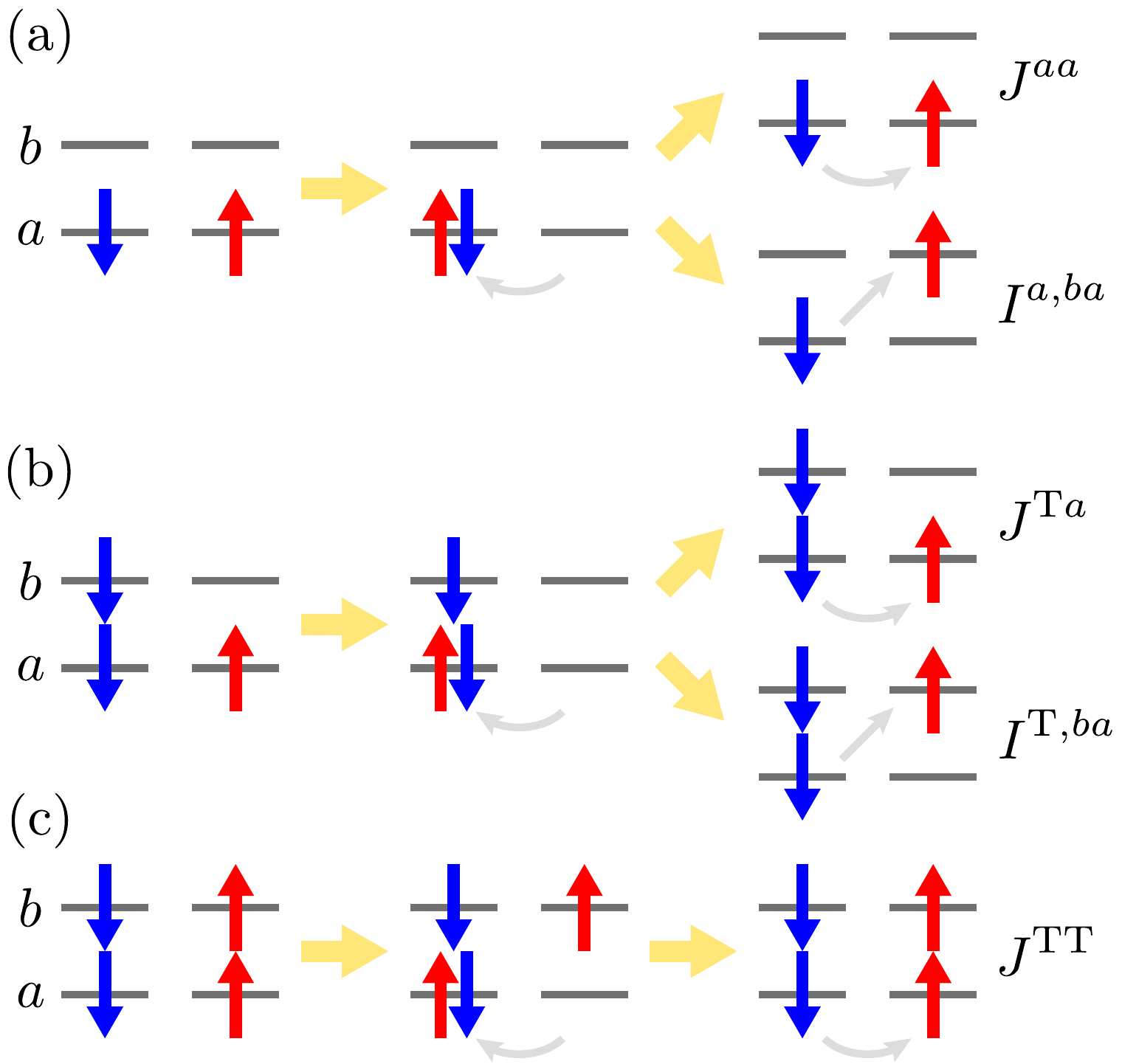} 
\caption{Second-order processes in the (a) spin-$\frac{1}{2}$ -- spin-$\frac{1}{2}$, (b) spin-$\frac{1}{2}$ -- spin-1, and (c) spin-1 -- spin-1 interactions.} 
\label{fig2}
\end{center}
\end{figure}

When one of the two sites is the singly occupied state ($\mu$) and the other is the spin-triplet state (T) as shown in Fig.~\ref{fig2}(b), the particle-exchange processes give the spin-$\frac{1}{2}$ -- spin-1 interactions. 
The $(\bm{R}',\nu) \rightarrow (\bm{R},\mu) \rightarrow (\bm{R}',\nu)$ process leads to the Hamiltonian 
\begin{align}
\hat{\mathcal{H}}^{(2)}_{J,\frac{1}{2}1}
= \sum_{\langle \bm{R},\bm{R}' \rangle} \sum_{\nu} & J^{{\rm T}\nu}_{\bm{R}\bm{R}'} 
\hat{P}_{\bm{R}}({\rm T}) \hat{P}_{\bm{R}'}(\nu)
\left( \hat{\bm{S}}_{\bm{R}} \cdot \hat{\bm{s}}_{\bm{R}',\nu} - \frac{1}{2} \right) 
\notag \\
& \times \hat{P}_{\bm{R}}({\rm T}) \hat{P}_{\bm{R}'}(\nu)
+\left( \bm{R} \leftrightarrow \bm{R}' \right) 
\end{align}
with 
\begin{align}
&J^{{\rm T}\nu}_{\bm{R}\bm{R}'} = \sum_{\mu} 
\left(  
\frac{|t^{\mu\nu}_{\bm{R}\bm{R}'}|^2}{U \!+\! U' \!+\! \Delta_{\mu\nu}}  
+ \frac{|t^{\mu\nu}_{\bm{R}\bm{R}'}|^2}{U \!-\! U' \!+\! J_{\rm H} \!-\! \Delta_{\mu\nu}} 
\right). 
\label{eq:J_Tnu}
\end{align} 
Additionally, the $(\bm{R}',\nu) \rightarrow (\bm{R},\mu) \rightarrow (\bm{R}',\bar{\nu})$ process [Fig.~\ref{fig2}(b)] leads to the correlated interorbital hopping term
\begin{align}
&\hat{\mathcal{H}}^{(2)}_{I,\frac{1}{2}1} 
= \sum_{\langle \bm{R},\bm{R}' \rangle} \sum_{\nu} 
I^{{\rm T}, \bar{\nu}\nu}_{\bm{R}\bm{R}'} 
\hat{P}_{\bm{R}}({\rm T}) \hat{P}_{\bm{R}'}(\bar{\nu}) 
\times
\notag \\
&\left[
\sum_{\sigma,\sigma'} \hat{c}^{\dag}_{\bm{R}',\bar{\nu},\sigma'}
\left( \hat{\bm{S}}_{\bm{R}} \! \cdot \!  \frac{\bm{\sigma}_{\sigma'\sigma}}{2} \! - \! \frac{1}{2} \delta_{\sigma',\sigma}\right) 
\hat{c}_{\bm{R}',\nu,\sigma} 
\right]
\hat{P}_{\bm{R}}({\rm T}) \hat{P}_{\bm{R}'}(\nu) 
\notag \\
& \qquad \qquad + \left( \bm{R} \leftrightarrow \bm{R}' \right) 
\label{eq:H_I_sT}
\end{align}
with 
\begin{align} 
&I^{{\rm T}, \bar{\nu}\nu}_{\bm{R} \bm{R}'} 
= \frac{1}{2} \sum_{\mu} 
\left( 
\frac{t^{\bar{\nu}\mu}_{\bm{R}'\bm{R}}t^{\mu\nu}_{\bm{R}\bm{R}'}}{U \!+\! U' \!+\! \Delta_{\mu\nu} } 
+ \frac{t^{\bar{\nu}\mu}_{\bm{R}'\bm{R}}t^{\mu\nu}_{\bm{R}\bm{R}'}}{U \!+\! U' \!+\! \Delta_{\mu\bar{\nu}}} 
\right). 
\label{eq:I_Tnunu}
\end{align}

\subsubsection{spin-$1$ -- spin-$1$ interaction}

When both $\bm{R}$ and $\bm{R}'$ sites are the spin-triplet states (T) [Fig.~\ref{fig2}(c)], the particle exchange processes lead to the spin-1 -- spin-1 interaction.   
Using the spin-1 operator $\hat{\bm{S}}_{\bm{R}}$, the Hamiltonian is given by 
\begin{align}
\hat{\mathcal{H}}^{(2)}_{J,11}
=\sum_{\langle \bm{R},\bm{R}' \rangle}
J^{{\rm TT}}_{\bm{R}\bm{R}'}
&\hat{P}_{\bm{R}}({\rm T}) \hat{P}_{\bm{R}'}({\rm T})  
\left( \hat{\bm{S}}_{\bm{R}} \cdot \hat{\bm{S}}_{\bm{R}'} - 1 \right)
\notag \\
&\times \hat{P}_{\bm{R}}({\rm T}) \hat{P}_{\bm{R}'}({\rm T}) 
\end{align}
with 
\begin{align}
&J^{\rm TT}_{\bm{R}\bm{R}'}
=\frac{|t^{aa}_{\bm{R}\bm{R}'}|^2 + |t^{bb}_{\bm{R}\bm{R}'}|^2}{U \!+\! J_{\rm H}} 
+\frac{ |t^{ab}_{\bm{R}\bm{R}'}|^2}{U \!+\! J_{\rm H} \!+\! \Delta} 
+\frac{ |t^{ab}_{\bm{R}\bm{R}'}|^2}{U \!+\! J_{\rm H} \!-\! \Delta}. 
\label{eq:J_TT}
\end{align}

\subsection{Effective model} 

The effective model, excluding doubly occupied orbitals, is described by 
\begin{align}
\hat{\mathcal{H}}_{\rm eff}  
= \hat{\mathcal{H}}_0  + \hat{\mathcal{T}}_0 + \hat{\mathcal{H}}^{(2)}_{\rm eff}. 
\label{eq:Heff_v1}
\end{align}
$\hat{\mathcal{H}}_0$ is the same as Eq.~(\ref{eq:H_TOHM}). 
Using the projection operators, $\hat{\mathcal{H}}_0 = \sum_{\bm{R}} \hat{\mathcal{H}}_{0;\bm{R}}$ is given by $\hat{\mathcal{H}}_{0;\bm{R}}=\Delta \hat{P}_{\bm{R}}(b) + \left( U' - J_{\rm H} + \Delta \right) \hat{P}_{\bm{R}}({\rm T}) + \left( U' + J_{\rm H} + \Delta \right)\hat{P}_{\bm{R}}({\rm S})$. 
$\hat{\mathcal{T}}_0$ is the hopping term given by Eq.~(\ref{eq:T0}) that conserves $N_d$. 
$\hat{\mathcal{H}}^{(2)}_{\rm eff}$ represents the effective interactions derived from the second-order perturbation theory. 
This effective model is composed of the nine single-site states, $0$, $\uparrow_a$, $\downarrow_a$, $\uparrow_b$, $\downarrow_b$, S, T$_{+}$, T$_0$, and T$_-$. 
Hence, the number of single-site states $d$ is reduced from $d=16$ in the two-orbital Hubbard model to $d=9$ in the effective model. 
When $J_{\rm H}$ is strong, we can omit the spin-singlet (S) state approximately, where we can further reduce the number of local states to $d=8$. 

We comment on the validity of the derived effective model. 
$J^{{\rm T}\nu}_{\bm{R}\bm{R}'}$ in Eq.~(\ref{eq:J_Tnu}) includes the spin-exchange term $|t^{\nu\nu}_{\bm{R}\bm{R}'}|^2 /(U - U' + J_{\rm H})$. 
However, this term diverges when $U=U'$ and $J_{\rm H}=0$. 
This is because a hopping process $\left(\genfrac{}{}{0pt}{}{\circ}{\downarrow}\right)_{\bm{R}} \left(\genfrac{}{}{0pt}{}{\uparrow}{\uparrow}\right)_{\bm{R}'}$ $\to$ $\left(\genfrac{}{}{0pt}{}{\circ}{\uparrow \! \downarrow}\right)_{\bm{R}} \left(\genfrac{}{}{0pt}{}{\uparrow}{\circ}\right)_{\bm{R}'}$ does not take an energy cost when $U=U'$ and $J_{\rm H}=0$, indicating that we cannot exclude doubly occupied orbitals in the model.  
In this case, we should use the original two-orbital Hubbard model. 
However, because $J_{\rm H}$ is usually positive and $U'$ is smaller than $U$, $|t^{\nu\nu}_{\bm{R}\bm{R}'}|^2 /(U - U' + J_{\rm H})$ becomes a finite value. 
If $U' = U -2J_{\rm H}$~\cite{CCastellani1978}, $|t^{\nu\nu}_{\bm{R}\bm{R}'}|^2 /(U - U' + J_{\rm H}) = |t^{\nu\nu}_{\bm{R}\bm{R}'}|^2 /(3J_{\rm H})$, indicating that the perturbatively obtained spin interaction is valid at $3J_{\rm H} \gg t^{\mu\nu}_{\bm{R}\bm{R}'}$. 
In this condition, omitting the S (spin-singlet) state from the effective model becomes a reasonable approximation. 
Note that if the pair-hopping term is considered in the two-orbital Hubbard model, $\Delta E(\mu,\xi')$ ($\xi'=\nu$, S, T) in Table~\ref{table2} (i.e., the formulas of $J^{\mu\nu}_{\bm{R}\bm{R}'}$, $J^{{\rm T}\nu}_{\bm{R}\bm{R}'}$ and $I^{\mu,\bar{\nu}\nu}_{\bm{R}\bm{R}'}$) are partially modified (see Appendix~\ref{appendix_A}).


\section{Exclusion of empty sites} \label{sec:model_II}

When the mean occupancy of a single site is larger than one, i.e., $\braket{\hat{n}_{\bm{R},a}}+\braket{\hat{n}_{\bm{R},b}} \ge 1$, the interorbital repulsion $U'$ favors the absence of the empty state (unoccupied $a$ and $b$ orbitals) because the presence of an empty site indicates the presence of an excess S or T site that increases the energy of $U'$ ($+\Delta \pm J_{\rm H}$). 
For example, the occupancy of the $d_{3z^2-r^2}$ and $d_{x^2-y^2}$ orbitals in the bilayer nickelate La$_3$Ni$_2$O$_7$ is considered as 1.5~\cite{HSun2023}, where the probability of an empty site appearing is low in the ground state. 
In this case, we can exclude empty sites from the effective system. 
Here, we derive an effective model that excludes empty sites from the ground-state configuration, whereas we incorporate the spin interactions obtained with the second-order perturbation theory.

\subsection{Second-order perturbation theory}

\begin{figure}[b]
\begin{center}
\includegraphics[width=0.8\columnwidth]{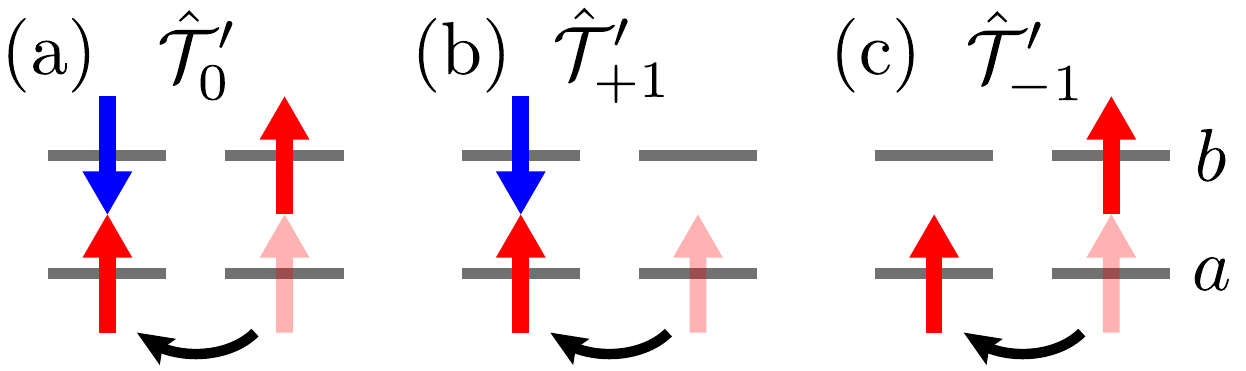} 
\caption{Hoppings in (a) $\hat{\mathcal{T}}'_0$, (b) $\hat{\mathcal{T}}'_{+1}$, and (c) $\hat{\mathcal{T}}'_{-1}$.} 
\label{fig3}
\end{center}
\end{figure}

The hopping term $\hat{\mathcal{T}}_0$ in Eq.~(\ref{eq:T0}) used in the previous section includes the processes creating an empty site due to $\hat{c}^{\dag}_{\bm{R},\mu,\sigma}\hat{c}_{\bm{R}',\nu,\sigma} \hat{P}_{\bm{R}'}(\nu)$, which annihilates an electron from the singly occupied site at $\bm{R}'$. 
To make an effective model without empty sites when $U' \gg t^{\mu\nu}_{\bm{R}\bm{R}'}$, we regroup the hopping terms in $\hat{\mathcal{T}}$.  
Here, we decompose the hopping term into
\begin{align}
\hat{\mathcal{T}} = \hat{\mathcal{T}}'_0 + \hat{\mathcal{T}}_{+1} + \hat{\mathcal{T}}'_{+1} + \hat{\mathcal{T}}_{-1} + \hat{\mathcal{T}}'_{-1}. 
\end{align}
$\hat{\mathcal{T}}_{+1}$ and $\hat{\mathcal{T}}_{-1}$ are the same as Eqs.(\ref{eq:T+}) and (\ref{eq:T-}), respectively. 
$\hat{\mathcal{T}}_0$ in Eq.~(\ref{eq:T0}) is divided into $\hat{\mathcal{T}}'_0$, $\hat{\mathcal{T}}'_{+1}$, and $\hat{\mathcal{T}}'_{-1}$, where 
\begin{align}
\hat{\mathcal{T}}'_0 
=-\sum_{\bm{R},\bm{R}'} \sum_{\mu,\nu} \sum_{\sigma} t^{\mu\nu}_{\bm{R}\bm{R}'}  
\Bigl[ 
\hat{P}_{\bm{R}}({\rm{S}}) + \hat{P}_{\bm{R}}({\rm{T}})
\Bigr] 
\notag \\
\times
\hat{c}^{\dag}_{\bm{R},\mu,\sigma} \hat{c}_{\bm{R}',\nu,\sigma}   
\Bigl[ 
\hat{P}_{\bm{R}'}({\rm{S}})
+\hat{P}_{\bm{R}'}({\rm{T}})
\Bigr], 
\label{eq:T'0}
\end{align}
\begin{align}
\hat{\mathcal{T}}'_{+1}
=-\sum_{\bm{R},\bm{R}'} \sum_{\mu,\nu} \sum_{\sigma} t^{\mu\nu}_{\bm{R}\bm{R}'} \left[ \hat{P}_{\bm{R}}({\rm{S}}) + \hat{P}_{\bm{R}}({\rm{T}}) \right] 
\notag \\
\times
\hat{c}^{\dag}_{\bm{R},\mu,\sigma} \hat{c}_{\bm{R}',\nu,\sigma}   
\hat{P}_{\bm{R}'}(\nu), 
\label{eq:T'+1}
\end{align}
and $\hat{\mathcal{T}}'_{-1} = \hat{\mathcal{T}}'^{\dag}_{+1}$. 
Figure~\ref{fig3} shows examples of hopping processes in $\hat{\mathcal{T}}'_0$, $\hat{\mathcal{T}}'_{+1}$, and $\hat{\mathcal{T}}'_{-1}$. 
The hopping term $\hat{\mathcal{T}}'_0$ does not change the number of doubly occupied orbitals and the number of empty sites [see Fig.~\ref{fig3}(a)].  
Note that the hopping term $\hat{P}_{\bm{R}}(\mu)\hat{c}^{\dag}_{\bm{R},\mu,\sigma}\hat{c}_{\bm{R}',\nu,\sigma} \hat{P}_{\bm{R}'}(\nu)$ is omitted in $\hat{\mathcal{T}}'_0$ because we configure the ground state without empty sites and $\hat{P}_{\bm{R}}(\mu)\hat{c}^{\dag}_{\bm{R},\mu,\sigma}\hat{c}_{\bm{R}',\nu,\sigma} \hat{P}_{\bm{R}'}(\nu)$ is allowed only when one of the two sites is empty.  
The hopping term $\hat{\mathcal{T}}'_{+1}$ ($\hat{\mathcal{T}}'_{-1}$) creates (annihilates) one S or T site at $\bm{R}$ and one empty site at $\bm{R}'$ [see Figs.~\ref{fig3}(b) and \ref{fig3}(c)]. 

The effective interactions due to the second-order processes (creating and annihilating an empty site) are derived by excluding $\hat{\mathcal{T}}'_{\pm 1}$ in the Schrieffer-Wolff transformation. 
The intermediate states in the second-order processes are different from those using $\hat{\mathcal{S}}^{(1)}_{\pm 1}$.
Here, the operator $\hat{\mathcal{S}}^{(1)}$ is extended as $\hat{\mathcal{S}}^{(1)}_{+1} + \hat{\mathcal{S}}'^{(1)}_{+1} + \hat{\mathcal{S}}^{(1)}_{-1} + \hat{\mathcal{S}}'^{(1)}_{-1}$, where we introduce $\hat{\mathcal{S}}'^{(1)}_{\pm 1}$ satisfying 
\begin{align}
\left[ \hat{\mathcal{H}}_0 , i\hat{\mathcal{S}}'^{(1)}_{\pm 1}  \right] = \hat{\mathcal{T}}'_{\pm 1}. 
\end{align}
We find that $i\hat{\mathcal{S}}'^{(1)}_{+ 1}$ is given by 
\begin{align}
&i\hat{\mathcal{S}}'^{(1)}_{+ 1} =
\notag \\
&-\sum_{\bm{R},\bm{R}'} \sum_{\mu,\nu} \sum_{\sigma} \frac{t^{\mu\nu}_{\bm{R}\bm{R}'}}{U' \!+\! J_{\rm H} \!+\! \Delta_{\mu\nu}}
\hat{P}_{\bm{R}}({\rm S}) 
\hat{c}^{\dag}_{\bm{R},\mu,\sigma} \hat{c}_{\bm{R}',\nu,\sigma}
\hat{P}_{\bm{R}'}(\nu) 
\notag \\
&-\sum_{\bm{R},\bm{R}'} \sum_{\mu,\nu} \sum_{\sigma} \frac{t^{\mu\nu}_{\bm{R}\bm{R}'}}{U'\!-\! J_{\rm H} \!+\!  \Delta_{\mu\nu}}
\hat{P}_{\bm{R}}({\rm T}) 
\hat{c}^{\dag}_{\bm{R},\mu,\sigma} \hat{c}_{\bm{R}',\nu,\sigma}
\hat{P}_{\bm{R}'}(\nu). 
\label{eq:SWT_S'1_+1}
\end{align}
Here, the S and T states are incorporated as the intermediate states in second-order processes. 
The energy changes are given by $U'+J_{\rm H}+\Delta_{\mu\nu}$ for the spin-singlet (S) creation and $U'-J_{\rm H}+\Delta_{\mu\nu}$ for the spin-triplet (T) creation.  
Then, the effective Hamiltonian is derived from  
\begin{align}
\hat{\mathcal{H}}'^{(2)}_{\rm eff} 
= -\frac{1}{2} \hat{\mathcal{T}}'_{-1} \left( i\hat{\mathcal{S}}'^{(1)}_{+1} \right) + {\rm H.c.}  
\end{align} 
The derivation of Eq.~(\ref{eq:SWT_S'1_+1}) and the details of $\hat{\mathcal{H}}'^{(2)}_{\rm eff} $ are given in Appendix~\ref{appendix_D}.

\subsection{Spin interactions} \label{sec:model_II_C}

The effective interactions obtained with second-order perturbation theory are described by $\hat{\mathcal{H}}'^{(2)}_{+}+\hat{\mathcal{H}}'^{(2)}_{-}$ with 
\begin{align}
\hat{\mathcal{H}}'^{(2)}_{\pm} 
&= \sum_{\langle \bm{R},\bm{R}' \rangle} \sum_{\mu,\mu',\nu,\nu'} 
K'^{\mu'\mu, \nu'\nu}_{\bm{R}\bm{R}';\pm} 
\hat{W}^{\mu'\mu, \nu'\nu}_{\bm{R}\bm{R}';\pm}. 
\end{align}
Here, the coupling constant $K'^{\mu'\mu,\nu'\nu}_{\bm{R}\bm{R}';\pm}$ and the operator $\hat{W}^{\mu'\mu,\nu'\nu}_{\bm{R}\bm{R}';\pm}$ are given by 
\begin{align}
K'^{\mu'\mu,\nu'\nu}_{\bm{R}\bm{R}';+} 
= \frac{1}{2} 
\Biggl( 
&\frac{t^{\nu'\bar{\mu}'}_{\bm{R}'\bm{R}} t^{\bar{\mu}\nu}_{\bm{R}\bm{R}'}}{U' + J_{\rm H} + \Delta_{\bar{\mu}\nu}}
+\frac{t^{\nu'\bar{\mu}'}_{\bm{R}'\bm{R}} t^{\bar{\mu}\nu}_{\bm{R}\bm{R}'}}{U' + J_{\rm H} + \Delta_{\bar{\mu}'\nu'}}
\notag \\
+&\frac{t^{\bar{\nu}\mu}_{\bm{R}'\bm{R}} t^{\mu'\bar{\nu}'}_{\bm{R}\bm{R}'}}{U' + J_{\rm H} + \Delta_{\bar{\nu}\mu}}
+\frac{t^{ \bar{\nu}\mu}_{\bm{R}'\bm{R}} t^{\mu'\bar{\nu}'}_{\bm{R}\bm{R}'}}{U' + J_{\rm H} + \Delta_{\bar{\nu}'\mu'}} 
\Biggr), 
\label{eq:int_KK_p} 
\end{align}
\begin{align}
\hat{W}^{\mu'\mu,\nu'\nu}_{\bm{R}\bm{R}';+} 
=&- \! \sum_{\sigma_1,\sigma_2,\sigma_3,\sigma_4}
\left(
\frac{1}{4}\delta_{\sigma_1,\sigma_2} \delta_{\sigma_3,\sigma_4} \!-\! \frac{1}{4} \bm{\sigma}_{\sigma_1\sigma_2} \cdot \bm{\sigma}_{\sigma_3\sigma_4} 
\right)
\notag \\
\times & \hat{P}_{\bm{R}}(\mu') \hat{P}_{\bm{R}'}(\nu') 
\hat{c}^{\dag}_{\bm{R},\mu',\sigma_1} \hat{c}_{\bm{R},\mu,\sigma_2}
\hat{c}^{\dag}_{\bm{R}',\nu',\sigma_3} \hat{c}_{\bm{R}',\nu,\sigma_4}
\notag \\
\times & \hat{P}_{\bm{R}}(\mu) \hat{P}_{\bm{R}'}(\nu) 
\end{align}
for the spin-singlet (S) intermediate state and 
\begin{align}
&K'^{\mu'\mu,\nu'\nu}_{\bm{R}\bm{R}';-}
\notag \\
&= \frac{\delta_{\mu',\mu} \!-\! \delta_{\mu',\bar{\mu}}}{2} 
\Biggl( 
\frac{t^{\nu'\bar{\mu}'}_{\bm{R}'\bm{R}} t^{\bar{\mu}\nu}_{\bm{R}\bm{R}'}}{U' - J_{\rm H} + \Delta_{\bar{\mu}\nu}}
\!+\! \frac{t^{\nu'\bar{\mu}'}_{\bm{R}'\bm{R}} t^{\bar{\mu}\nu}_{\bm{R}\bm{R}'}}{U' - J_{\rm H} + \Delta_{\bar{\mu}'\nu'}}
\Biggr)
\notag \\
&+\frac{\delta_{\nu',\nu} \!-\! \delta_{\nu',\bar{\nu}}}{2} 
\Biggl( 
\frac{t^{\bar{\nu}\mu}_{\bm{R}'\bm{R}} t^{\mu'\bar{\nu}'}_{\bm{R}\bm{R}'}}{U' - J_{\rm H} + \Delta_{\bar{\nu}\mu}}
\!+\! \frac{t^{ \bar{\nu} \mu}_{\bm{R}'\bm{R}} t^{\mu'\bar{\nu}'}_{\bm{R}\bm{R}'}}{U' - J_{\rm H} + \Delta_{\bar{\nu}'\mu'}}
\Biggr), 
\label{eq:int_KK_m} 
\end{align}
\begin{align}
\hat{W}^{\mu'\mu, \nu'\nu}_{\bm{R}\bm{R}';-} =
&- \! \sum_{\sigma_1,\sigma_2,\sigma_3,\sigma_4}  
\left(
\frac{3}{4}\delta_{\sigma_1,\sigma_2} \delta_{\sigma_3,\sigma_4} 
\!+\! \frac{1}{4} \bm{\sigma}_{\sigma_1\sigma_2} \cdot \bm{\sigma}_{\sigma_3\sigma_4} 
\right) 
\notag \\
\times & \hat{P}_{\bm{R}}(\mu') \hat{P}_{\bm{R}'}(\nu') 
\hat{c}^{\dag}_{\bm{R},\mu',\sigma_1} \hat{c}_{\bm{R},\mu,\sigma_2}  
\hat{c}^{\dag}_{\bm{R}',\nu',\sigma_3} \hat{c}_{\bm{R}',\nu,\sigma_4}  
\notag \\
\times & \hat{P}_{\bm{R}}(\mu) \hat{P}_{\bm{R}'}(\nu)  
\end{align}
for the spin-triplet (T) intermediate state.   

\begin{figure}[t]
\begin{center}
\includegraphics[width=\columnwidth]{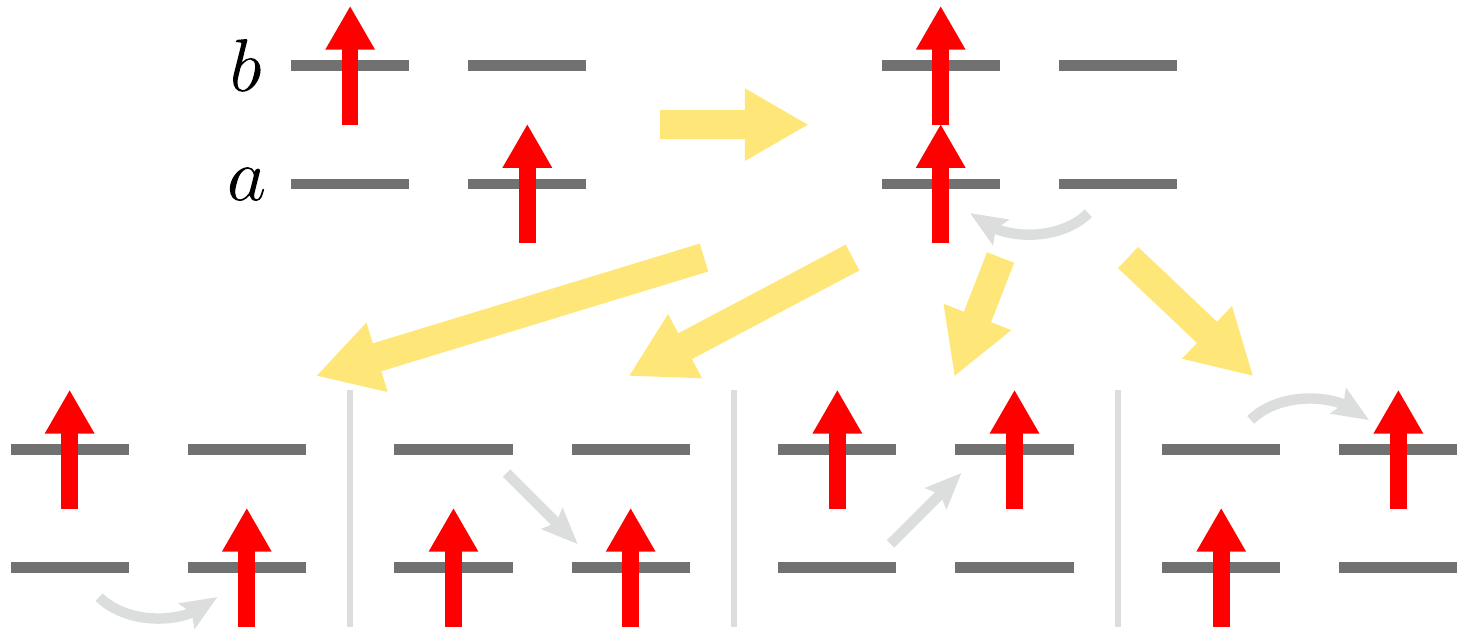} 
\caption{Second-order processes in $\hat{\mathcal{H}}'^{(2)}_{\rm eff}$.} 
\label{fig4}
\end{center}
\end{figure}

Examples of possible second-order processes are shown in Fig.~\ref{fig4}. 
If we introduce the pseudospin operator for the orbital degree of freedom, $\hat{\mathcal{H}}'^{(2)}_{\pm}$ can be written as Kugel-Khomskii--type coupling~\cite{KKugel1982,SIshihara1997}. 
Note that $\hat{\mathcal{H}}'^{(2)}_{-}$ gives the ferromagnetic spin interaction. 
In particular, when $\mu'=\mu$ and $\nu'=\nu$,  
\begin{align}
&K'^{\mu\mu,\nu\nu}_{\bm{R}\bm{R}';\pm} = 
\frac{t^{\nu\bar{\mu}}_{\bm{R}'\bm{R}} t^{\bar{\mu}\nu}_{\bm{R}\bm{R}'}}{U' \pm J_{\rm H} + \Delta_{\bar{\mu}\nu}}
+\frac{t^{\mu\bar{\nu}}_{\bm{R}\bm{R}'} t^{\bar{\nu}\mu}_{\bm{R}'\bm{R}}}{U' \pm J_{\rm H} + \Delta_{\bar{\nu}\mu}}, 
\end{align}
and 
\begin{align}
&\hat{W}^{\mu\mu,\nu\nu}_{\bm{R}\bm{R}';+} 
\!=\! \hat{P}_{\bm{R}}(\mu) \hat{P}_{\bm{R}'}(\nu) 
\!\left(
\hat{\bm{s}}_{\bm{R},\mu} \! \cdot \! \hat{\bm{s}}_{\bm{R}',\nu} 
- \frac{1}{4} 
\right) \!
\hat{P}_{\bm{R}}(\mu) \hat{P}_{\bm{R}'}(\nu) ,
\\
&\hat{W}^{\mu\mu,\nu\nu}_{\bm{R}\bm{R}';-} 
\!=\! - \hat{P}_{\bm{R}}(\mu) \hat{P}_{\bm{R}'}(\nu) 
\! \left(
\hat{\bm{s}}_{\bm{R},\mu} \! \cdot \! \hat{\bm{s}}_{\bm{R}',\nu} 
+\frac{3}{4} 
\right) \! 
\hat{P}_{\bm{R}}(\mu) \hat{P}_{\bm{R}'}(\nu) . 
\end{align}
Since $K'^{\mu\mu,\nu\nu}_{\bm{R}\bm{R}';-} > K'^{\mu\mu,\nu\nu}_{\bm{R}\bm{R}';+}$ when $J_{\rm H}>0$, the ferromagnetic spin correlation is favorably generated from these spin interactions.  

\subsection{Effective model} 

The effective model that excludes doubly occupied orbitals and empty sites is described by 
\begin{align}
\hat{\mathcal{H}}_{\rm eff}  
=\hat{\mathcal{H}}_0 + \hat{\mathcal{T}}'_0 + \hat{\mathcal{H}}^{(2)}_{\rm eff} + \hat{\mathcal{H}}'^{(2)}_{\rm eff}. 
\end{align}
$\hat{\mathcal{H}}_0$ and $\hat{\mathcal{H}}^{(2)}_{\rm eff}$ are the same as the Hamiltonians in Eq.~(\ref{eq:Heff_v1}). 
Because the hopping term that creates an empty site is excluded, $\hat{\mathcal{T}}_0$ in Eq.~(\ref{eq:Heff_v1}) is replaced by $\hat{\mathcal{T}}'_0$ in Eq.~(\ref{eq:T'0}). 
$\hat{\mathcal{H}}'^{(2)}_{\rm eff}$ emerges due to the exclusion of empty sites. 
This effective model is composed of the eight single-site states, $\uparrow_a$, $\downarrow_a$, $\uparrow_b$, $\downarrow_b$, S, T$_{+}$, T$_0$, and T$_-$. 
When $J_{\rm H}$ is sufficiently large, we can omit the spin-singlet (S) state. 
Hence, when Hund's coupling $J_{\rm H}$ is active, we can reduce the number of single-site states to $d=7$.


\section{Effective model for bilayer nickelate superconductors} \label{sec:model_LNO}

\begin{figure}[b]
\begin{center}
\includegraphics[width=\columnwidth]{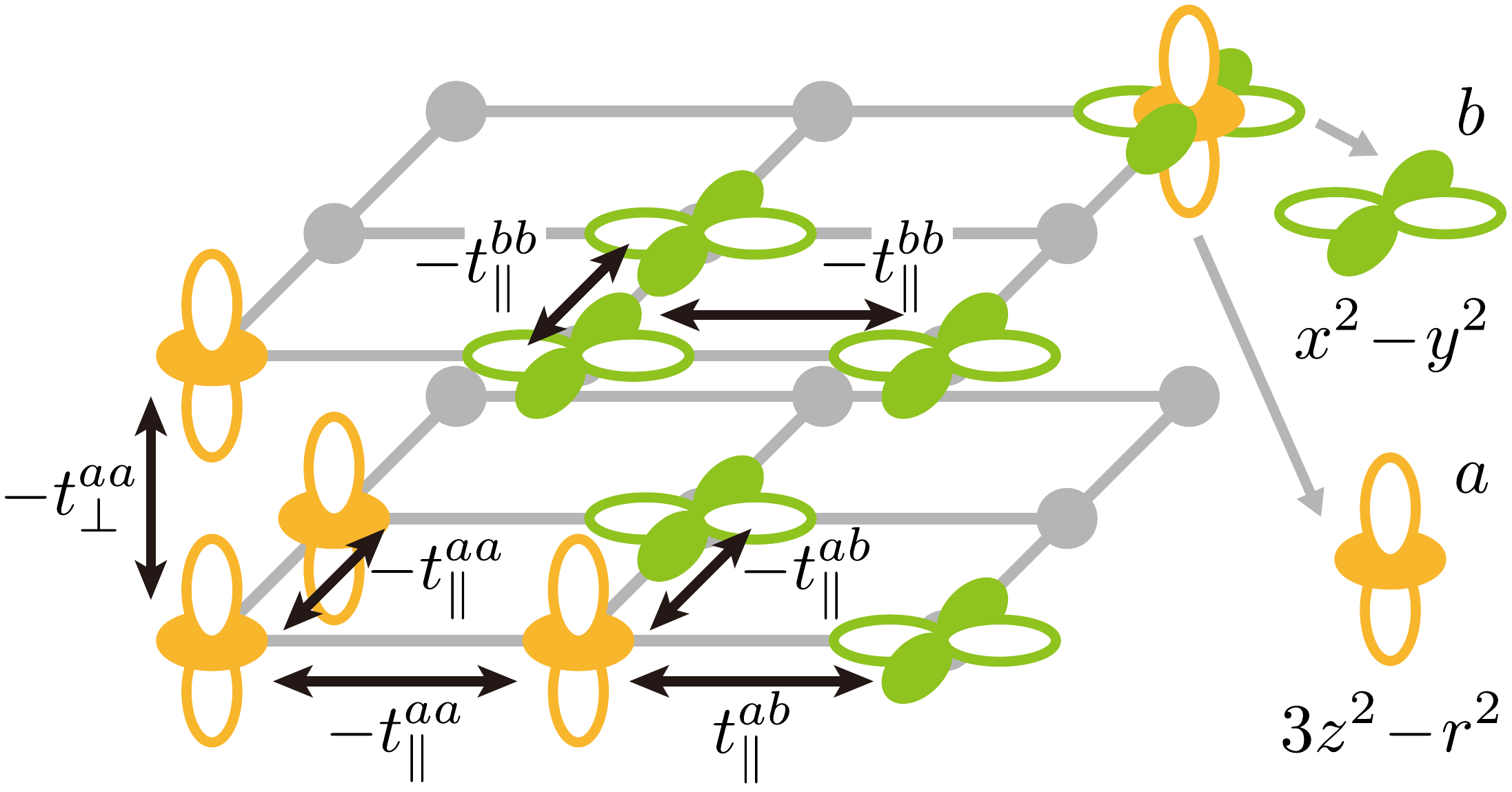}
\caption{Two-orbital bilayer system for the bilayer nickelate La$_3$Ni$_2$O$_7$, where $a$ and $b$ represent the $d_{3z^2-r^2}$ and $d_{x^2-y^2}$ orbitals, respectively.} 
\label{fig5}
\end{center}
\end{figure}

We present an effective model adapted to bilayer nickelate superconductors. 
In the bilayer nickelate La$_3$Ni$_2$O$_7$, the electron configuration of Ni is considered as $d^{7.5}$~\cite{HSun2023}, where the $d_{3z^2-r^2}$ and $d_{x^2-y^2}$ orbitals mainly determine the electronic properties. 
The two-orbital Hubbard model in bilayer geometry shown in Fig.~\ref{fig5} can capture the correlated $d$ electrons in bilayer nickelate superconductors~\cite{ZLuo2023,HSakakibara2024_327}. 
Due to the crystal-field splitting, the energy level of the $d_{3z^2-r^2}$ orbital is lower than that of the $d_{x^2-y^2}$ orbital.  
Hence, the $d_{3z^2-r^2}$ orbital is nearly half-filling, while the $d_{x^2-y^2}$ orbital is nearly quarter-filling. 
In this section, we set these two orbitals as $a = d_{3z^2-r^2}$ and $b = d_{x^2-y^2}$, where the mean occupancy is $\braket{\hat{n}_{\bm{R},a}}+\braket{\hat{n}_{\bm{R},b}} = 1.5$.  
Since the occupancy is larger than one, here, we consider the effective model without empty sites and doubly occupied orbitals. 
We introduce the model that omits the S (spin-singlet) state, assuming that $J_{\rm H}$ is sufficiently large. 
Since the bilayer system comprises the interlayer and intralayer hoppings, we present the resulting interlayer and intralayer spin interactions obtained with the second-order perturbation theory. 

In this section, for notational simplicity, the spin operators sandwiched by the projection operators are expressed as 
\begin{align}
&\hat{\tilde{\bm{s}}}_{\bm{R},\mu} = 
\hat{P}_{\bm{R}}(\mu) \hat{\bm{s}}_{\bm{R},\mu} \hat{P}_{\bm{R}}(\mu) , 
\\
&\hat{\tilde{\bm{S}}}_{\bm{R}} = 
\hat{P}_{\bm{R}}({\rm T}) \hat{\bm{S}}_{\bm{R}} \hat{P}_{\bm{R}}(\rm{T}) , 
\end{align}
and we redefine the projection operator as 
\begin{align}
&\hat{\tilde{n}}_{\bm{R},\xi} = \hat{P}_{\bm{R}}(\xi)  . 
\end{align}
To describe the onsite interorbital hopping, we introduce 
\begin{align}
&\hat{\tilde{\rho}}_{\bm{R},\bar{\mu}\mu} = 
\hat{P}_{\bm{R}}(\bar{\mu}) 
\left[
\sum_{\sigma}
\hat{c}^{\dag}_{\bm{R},\bar{\mu},\sigma}  
\hat{c}_{\bm{R},\mu,\sigma}  
\right]
\hat{P}_{\bm{R}}(\mu) 
\end{align}
and 
\begin{align}
&\hat{\tilde{\bm{\tau}}}_{\bm{R},\bar{\mu}\mu} = 
\hat{P}_{\bm{R}}(\bar{\mu}) 
\left[\frac{1}{2}
\sum_{\sigma_1,\sigma_2}
\hat{c}^{\dag}_{\bm{R},\bar{\mu},\sigma_1}  
\bm{\sigma}_{\sigma_1\sigma_2}
\hat{c}_{\bm{R},\mu,\sigma_2}  
\right]
\hat{P}_{\bm{R}}(\mu) . 
\end{align}
From the next subsection, the site at position $\bm{R}$ is indicated by $(j,l)$, where $j$ is the site index in the two-dimensional plane and $l=1,2$ is the layer index in the bilayer structure.

\subsection{Interlayer interactions}

First, we consider the interlayer spin interactions. 
In the non-tilted La$_3$Ni$_2$O$_7$ structure in the high-pressure phase that shows SC~\cite{HSun2023}, the interlayer hopping $t^{ab}_{\perp}$ between the $d_{3z^2-r^2}$ ($a$) and $d_{x^2-y^2}$ ($b$) orbitals is zero. 
Because the $d_{x^2-y^2}$ orbital is elongated in the in-plane direction (see Fig.~\ref{fig5}), the interlayer hopping $t^{bb}_{\perp}$ between the $d_{x^2-y^2}$ ($b$) orbitals is tiny~\cite{ZLuo2023}. 
Hence, we only consider the interlayer hopping $t^{aa}_{\perp}$ between the $d_{3z^2-r^2}$ ($a$) orbitals, i.e., $t^{aa}_{\perp} \ne 0$ and $t^{bb}_{\perp} = t^{ab}_{\perp} = 0$. 

Since $t^{ab}_{\perp}=0$, the number of effective interactions is limited. 
In particular, the correlated hopping $I^{\xi,\bar{\nu}\nu}_{\perp}$ is zero due to $t^{ab}_{\perp}=0$. 
The two-site spin interactions in $\hat{\mathcal{H}}^{(2)}_{{\rm eff}}$ are given by 
\begin{align}
\hat{\mathcal{H}}^{(2)}_{{\rm eff}; \perp}
&=J^{aa}_{\perp} \sum_{j} 
\left( 
\hat{\tilde{\bm{s}}}_{j,1,a} \cdot \hat{\tilde{\bm{s}}}_{j,2,a} 
-\frac{1}{4} \hat{\tilde{n}}_{j,1,a}\hat{\tilde{n}}_{j,2,a}
\right) 
\notag \\
&+J^{{\rm T}a}_{\perp} 
\sum_{j} 
\left( 
\hat{\tilde{\bm{S}}}_{j,1} \cdot \hat{\tilde{\bm{s}}}_{j,2,a} 
-\frac{1}{2} \hat{\tilde{n}}_{j,1,{\rm T}}\hat{\tilde{n}}_{j,2,a}
\right) 
\notag \\
&+J^{a{\rm T}}_{\perp}
\sum_{j} 
\left( 
\hat{\tilde{\bm{s}}}_{j,1,a} \cdot \hat{\tilde{\bm{S}}}_{j,2} 
-\frac{1}{2} \hat{\tilde{n}}_{j,1,a}\hat{\tilde{n}}_{j,2,{\rm T}} 
\right) 
\notag \\
&+J^{{\rm TT}}_{\perp}
\sum_{j}
\left( 
\hat{\tilde{\bm{S}}}_{j,1} \cdot \hat{\tilde{\bm{S}}}_{j,2} 
-\hat{\tilde{n}}_{j,1,{\rm T}}\hat{\tilde{n}}_{j,2,{\rm T}} 
\right)
\label{eq:Heff_perp}
\end{align}
with $J^{aa}_{\perp} = 4|t^{aa}_{\perp}|^2 / U$, $J^{{\rm T}a}_{\perp}  = J^{a{\rm T}}_{\perp} = |t^{aa}_{\perp}|^2 / (U + U') + |t^{aa}_{\perp}|^2 / (U - U' + J_{\rm H})$, and $J^{\rm TT}_{\perp} = |t^{aa}_{\perp}|^2 / ( U + J_{\rm H} )$. 
In addition, the spin interactions in $\hat{\mathcal{H}}'^{(2)}_{\rm eff}$ lead to 
\begin{align}
\hat{\mathcal{H}}'^{(2)}_{{\rm eff};\perp} 
&=J'^{ab}_{\perp;+} \sum_{j} \sum_{\mu}
\left(
\hat{\tilde{\bm{s}}}_{j,1,\mu} \cdot \hat{\tilde{\bm{s}}}_{j,2,\bar{\mu}} 
- \frac{1}{4} \hat{\tilde{n}}_{j,1,\mu} \hat{\tilde{n}}_{j,2,\bar{\mu}} 
\right) 
\notag \\
&-J'^{ab}_{\perp;-} \sum_{j} \sum_{\mu}
\left(
\hat{\tilde{\bm{s}}}_{j,1,\mu} \cdot \hat{\tilde{\bm{s}}}_{j,2,\bar{\mu}} 
+ \frac{3}{4} \hat{\tilde{n}}_{j,1,\mu} \hat{\tilde{n}}_{j,2,\bar{\mu}} 
\right), 
\label{eq:Heff_perp_KK}
\end{align}
where $J'^{ab}_{\perp;\pm} = |t^{aa}_{\perp}|^2 / (U' \pm J_{\rm H} )$ is given by $ K'^{aa,bb}_{\bm{R}\bm{R}';\pm}$ in Eqs.~(\ref{eq:int_KK_p}) and (\ref{eq:int_KK_m}). 

\begin{figure}[b]
\begin{center}
\includegraphics[width=\columnwidth]{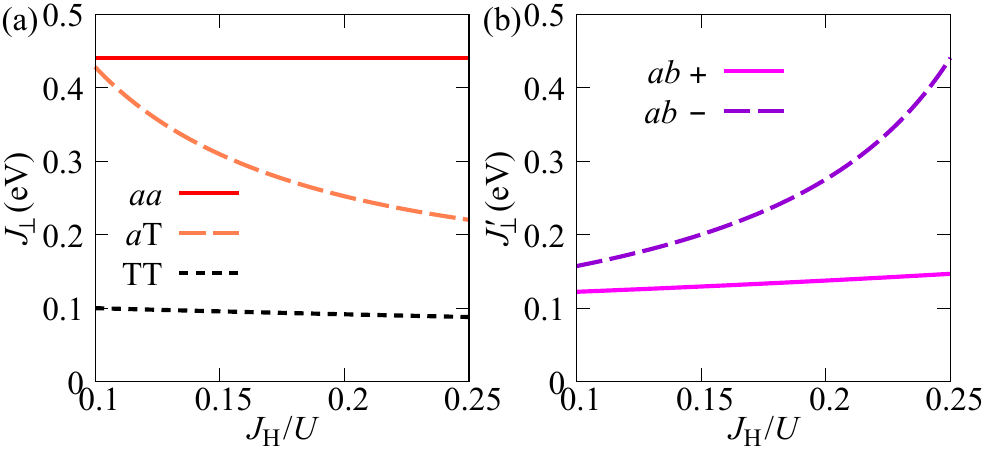} 
\caption{$J_{\rm H}$ dependencies of (a) $J^{aa}_{\perp}$, $J^{a{\rm T}}_{\perp}$, $J^{\rm TT}_{\perp}$, and (b) $J'^{ab}_{\perp;\pm}$, where $t^{aa}_{\perp}=0.664$~eV, $U=4$~eV, and $U' = U -2J_{\rm H}$.} 
\label{fig6}
\end{center}
\end{figure}

Figure~\ref{fig6} shows $J^{\xi\xi'}_{\perp}$ and $J'^{ab}_{\perp;\pm}$ depending on $J_{\rm H}$. 
Here, we use $U' = U -2J_{\rm H}$ with $U=4$~eV and employ the hopping parameter $t^{aa}_{\perp}=0.664$~eV evaluated by the first-principles calculation for La$_3$Ni$_2$O$_7$ in Ref.~\cite{HSakakibara2024_327}. 
As shown in Fig.~\ref{fig6}(a), the values of $J^{aa}_{\perp}$, $J^{a{\rm T}}_{\perp}$, and $J^{\rm TT}_{\perp}$ are of about the same order of magnitude. 
Since the size of spin 1 is twice as large as the size of spin $\frac{1}{2}$, $J^{aa}_{\perp}$ should be compared with $2J^{a{\rm T}}_{\perp}$ and $4J^{\rm TT}_{\perp}$. 
All $J^{\xi\xi'}_{\perp}$ work antiferromagnetically, whereas $J'^{ab}_{\perp;-} > J'^{ab}_{\perp;+}$ due to Hund's coupling $J_{\rm H}$ as shown in Fig.~\ref{fig6}(b), indicating that the ferromagnetic interaction does work between the interorbital spin-$\frac{1}{2}$ states.

\subsection{Intralayer interactions}

Next, we consider the intralayer interactions. 
In contrast to the interlayer hopping $t^{\mu\nu}_{\perp}$, the intraorbital ($aa$, $bb$) and interorbital ($ab$) hopping parameters have non-negligible values~\cite{ZLuo2023,HSakakibara2024_327}.  
Because the $d_{x^2-y^2}$ ($b$) orbital spreads in the in-plane ($xy$) direction while the $d_{3z^2-r^2}$ ($a$) orbital is elongated in the out-of-plane ($z$) direction (see Fig.~\ref{fig5}), the hopping parameters satisfy $|t^{bb}_{\parallel}| > |t^{ab}_{\parallel}| > |t^{aa}_{\parallel}|$, i.e., the intralayer hopping between the $d_{x^2-y^2}$ orbitals is the largest. 
While $t^{ab}_{\parallel}=t^{ba}_{\parallel}$, the signs of the interorbital hopping $t^{ab}_{\parallel}$ in the $x$ direction and $y$ direction are opposite as shown in Fig.~\ref{fig5}~\cite{ZLuo2023,YZhang2023}. 

In contrast to the interlayer interactions, the interorbital hopping $t^{ab}_{\parallel}$ allows the interaction $I^{\xi,\bar{\nu}\nu}_{\parallel}$. 
When all nearest-neighbor intralayer hopping parameters are considered, the in-plane two-site interactions in $\hat{\mathcal{H}}^{(2)}_{{\rm eff}}$ are given by 
\begin{align}
&\hat{\mathcal{H}}^{(2)}_{{\rm eff}; \parallel}
= \sum_{\mu,\nu} J^{\mu\nu}_{\parallel} \sum_{\langle i,j \rangle} \sum_{l}
\left( 
\hat{\tilde{\bm{s}}}_{i,l,\mu} \cdot \hat{\tilde{\bm{s}}}_{j,l,\nu} 
- \frac{1}{4} \hat{\tilde{n}}_{i,l,\mu}\hat{\tilde{n}}_{j,l,\nu}
\right) 
\notag \\
&+ \sum_{\nu} J^{{\rm T} \nu}_{\parallel} 
\sum_{\langle i,j \rangle} \sum_{l}
\left( 
\hat{\tilde{\bm{S}}}_{i,l} \cdot \hat{\tilde{\bm{s}}}_{j,l,\nu} 
- \frac{1}{2} \hat{\tilde{n}}_{i,l,{\rm T}}\hat{\tilde{n}}_{j,l,\nu}
\right) 
\notag \\
&+ \sum_{\mu} J^{\mu {\rm T}}_{\parallel} 
\sum_{\langle i,j \rangle} \sum_{l}
\left( 
\hat{\tilde{\bm{s}}}_{i,l,\mu} \cdot \hat{\tilde{\bm{S}}}_{j,l} 
- \frac{1}{2} \hat{\tilde{n}}_{i,l,\mu}\hat{\tilde{n}}_{j,l,{\rm T}} 
\right) 
\notag \\
&+J^{{\rm TT}}_{\parallel} 
\sum_{\langle i,j \rangle} \sum_{l}
\left( 
\hat{\tilde{\bm{S}}}_{i,l} \cdot \hat{\tilde{\bm{S}}}_{j,l} 
- \hat{\tilde{n}}_{i,l,{\rm T}}\hat{\tilde{n}}_{j,l,{\rm T}} \right)
\notag \\
&+ \sum_{\mu,\nu} 
I^{\mu, \bar{\nu}\nu}_{\parallel} 
\sum_{\langle i,j \rangle} \sum_{l} \gamma_{ij}
\left( 
\hat{\tilde{\bm{s}}}_{i,l,\mu} \! \cdot \! \hat{\tilde{\bm{\tau}}}_{j,l,\bar{\nu}\nu}
- \frac{1}{4} \hat{\tilde{n}}_{i,l,\mu}\hat{\tilde{\rho}}_{j,l,\bar{\nu}\nu}
\right)
\notag \\
&+ \sum_{\mu,\nu} 
I^{\bar{\mu} \mu, \nu}_{\parallel}
\sum_{\langle i,j \rangle} \sum_{l} \gamma_{ij}
\left( 
\hat{\tilde{\bm{\tau}}}_{i,l,\bar{\mu}\mu} \! \cdot \! \hat{\tilde{\bm{s}}}_{j,l,\nu}
- \frac{1}{4} \hat{\tilde{\rho}}_{i,l,\bar{\mu}\mu}\hat{\tilde{n}}_{j,l,\nu}
\right)
\notag \\
&+ \sum_{\nu} 
I^{{\rm T}, \bar{\nu}\nu}_{\parallel} 
\sum_{\langle i,j \rangle} \sum_{l} \gamma_{ij}
\left( 
\hat{\tilde{\bm{S}}}_{i,l} \! \cdot \! \hat{\tilde{\bm{\tau}}}_{j,l,\bar{\nu}\nu}
- \frac{1}{2} \hat{\tilde{n}}_{i,l,{\rm T}}\hat{\tilde{\rho}}_{j,l,\bar{\nu}\nu}
\right)
\notag \\
&+ \sum_{\mu} 
I^{\bar{\mu}\mu, {\rm T}}_{\parallel} 
\sum_{\langle i,j \rangle} \sum_{l} \gamma_{ij}
\left( 
\hat{\tilde{\bm{\tau}}}_{i,l,\bar{\mu}\mu} \! \cdot \! \hat{\tilde{\bm{S}}}_{j,l}
- \frac{1}{2} \hat{\tilde{\rho}}_{i,l,\bar{\mu}\mu}\hat{\tilde{n}}_{j,l,{\rm T}}
\right). 
\label{eq:Heff_para}
\end{align}
The coupling constants $J^{\xi\xi'}_{\parallel}$ and $I^{\xi,\bar{\nu}\nu}_{\parallel}$ are given by Eqs.~(\ref{eq:J_munu}), (\ref{eq:I_mununu}), (\ref{eq:J_Tnu}), (\ref{eq:I_Tnunu}), and (\ref{eq:J_TT}). 
The explicit formulas of $J^{\xi\xi'}_{\parallel}$ and $I^{\xi,\bar{\nu}\nu}_{\parallel}$ using $t^{aa}_{\parallel}$, $t^{bb}_{\parallel}$, and $t^{ab}_{\parallel}$ are presented in Appendix~\ref{appendix_E}. 
$\gamma_{ij}$ ($=\pm 1$) in the $I_{\parallel}$ terms is the sign depending on the direction of bond $i$-$j$ due to the sign of the interorbital hopping $t^{ab}_{\parallel}$. 
Additionally, the Hamiltonian of $\hat{\mathcal{H}}'^{(2)}_{\rm eff}$ is given by 
\begin{align}
&\hat{\mathcal{H}}'^{(2)}_{{\rm eff}; \parallel}
=\sum_{\mu,\nu} J'^{\mu\nu}_{\parallel;+} \sum_{\langle i,j \rangle} \sum_{l}
\left( 
\hat{\tilde{\bm{s}}}_{i,l,\mu} \cdot \hat{\tilde{\bm{s}}}_{j,l,\nu} 
-\frac{1}{4} \hat{\tilde{n}}_{i,l,\mu}\hat{\tilde{n}}_{j,l,\nu}
\right) 
\notag \\
&+\sum_{\mu,\nu} 
I'^{\mu , \bar{\nu}\nu}_{\parallel;+} 
\sum_{\langle i,j \rangle} \sum_{l} \gamma_{ij}
\left( 
\hat{\tilde{\bm{s}}}_{i,l,\mu} \! \cdot \! \hat{\tilde{\bm{\tau}}}_{j,l,\bar{\nu}\nu}
- \frac{1}{4} \hat{\tilde{n}}_{i,l,\mu}\hat{\tilde{\rho}}_{j,l,\bar{\nu}\nu}
\right)
\notag \\
&+\sum_{\mu,\nu} 
I'^{\bar{\mu}\mu, \nu}_{\parallel;+} 
\sum_{\langle i,j \rangle} \sum_{l} \gamma_{ij}
\left( 
\hat{\tilde{\bm{\tau}}}_{i,l,\bar{\mu}\mu} \! \cdot \! \hat{\tilde{\bm{s}}}_{j,l,\nu}
- \frac{1}{4} \hat{\tilde{\rho}}_{i,l,\bar{\mu}\mu}\hat{\tilde{n}}_{j,l,\nu}
\right)
\notag \\
&+\sum_{\mu,\nu} 
K'^{\bar{\mu} \mu, \bar{\nu}\nu}_{\parallel;+} 
\sum_{\langle i,j \rangle} \sum_{l}
\left( 
\hat{\tilde{\bm{\tau}}}_{i,l,\bar{\mu}\mu} \! \cdot \!  \hat{\tilde{\bm{\tau}}}_{j,l,\bar{\nu}\nu}
- \frac{1}{4} \hat{\tilde{\rho}}_{i,l,\bar{\mu}\mu} \hat{\tilde{\rho}}_{j,l,\bar{\nu}\nu}
\right)
\notag \\
&-\sum_{\mu,\nu} J'^{\mu\nu}_{\parallel;-} \sum_{\langle i,j \rangle} \sum_{l}
\left( 
\hat{\tilde{\bm{s}}}_{i,l,\mu} \cdot \hat{\tilde{\bm{s}}}_{j,l,\nu} 
+\frac{3}{4} \hat{\tilde{n}}_{i,l,\mu}\hat{\tilde{n}}_{j,l,\nu}
\right) 
\notag \\
&-\sum_{\mu,\nu} 
I'^{\mu, \bar{\nu}\nu}_{\parallel;-} 
\sum_{\langle i,j \rangle} \sum_{l} \gamma_{ij}
\left( 
\hat{\tilde{\bm{s}}}_{i,l,\mu} \! \cdot \! \hat{\tilde{\bm{\tau}}}_{j,l,\bar{\nu}\nu}
+\frac{3}{4} \hat{\tilde{n}}_{i,l,\mu}\hat{\tilde{\rho}}_{j,l,\bar{\nu}\nu}
\right)
\notag \\
&-\sum_{\mu,\nu} 
I'^{\bar{\mu}\mu, \nu}_{\parallel;-} 
\sum_{\langle i,j \rangle} \sum_{l} \gamma_{ij}
\left( 
\hat{\tilde{\bm{\tau}}}_{i,l,\bar{\mu}\mu} \! \cdot \! \hat{\tilde{\bm{s}}}_{j,l,\nu}
+\frac{3}{4} \hat{\tilde{\rho}}_{i,l,\bar{\mu}\mu}\hat{\tilde{n}}_{j,l,\nu}
\right)
\notag \\
&-\sum_{\mu,\nu} 
K'^{\bar{\mu} \mu, \bar{\nu}\nu}_{\parallel;-} 
\sum_{\langle i,j \rangle} \sum_{l} 
\left( 
\hat{\tilde{\bm{\tau}}}_{i,l,\bar{\mu}\mu} \! \cdot \! \hat{\tilde{\bm{\tau}}}_{j,l,\bar{\nu}\nu}
+\frac{3}{4} \hat{\tilde{\rho}}_{i,l,\bar{\mu}\mu} \hat{\tilde{\rho}}_{j,l,\bar{\nu}\nu}
\right). 
\label{eq:Heff_para_KK}
\end{align}
$J'^{\mu\nu}_{\parallel;\pm}$ and $I'^{\mu,\bar{\nu}\nu}_{\parallel;\pm}$ are given by $ K'^{\mu\mu,\nu\nu}_{\bm{R}\bm{R}';\pm}$ and $ K'^{\mu\mu, \bar{\nu}\nu}_{\bm{R}\bm{R}';\pm}$, respectively, while $K'^{\bar{\mu}\mu,\bar{\nu}\nu}_{\parallel;\pm}$ is the same as $K'^{\bar{\mu}\mu,\bar{\nu}\nu}_{\bm{R}\bm{R}';\pm}$ in Eqs.~(\ref{eq:int_KK_p}) and (\ref{eq:int_KK_m}). 
Their explicit formulas are presented in Appendix~\ref{appendix_E}. 

\begin{figure}[t]
\begin{center}
\includegraphics[width=\columnwidth]{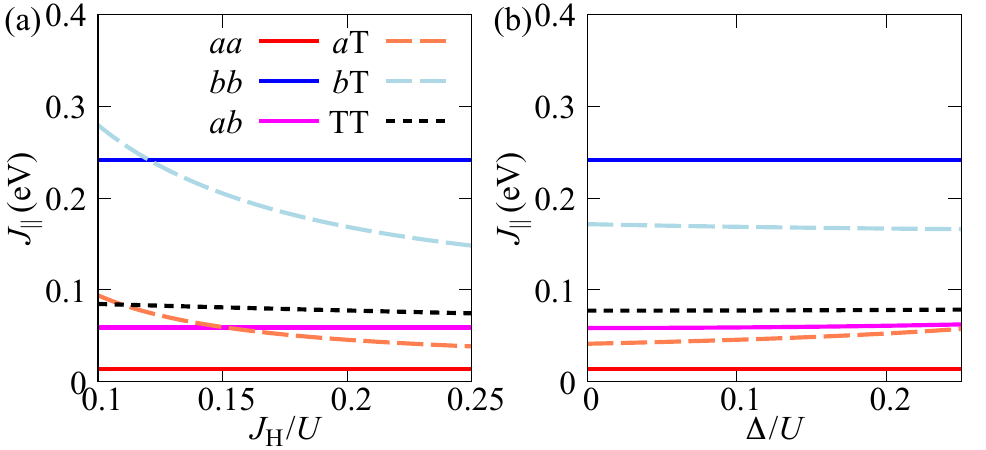} 
\includegraphics[width=\columnwidth]{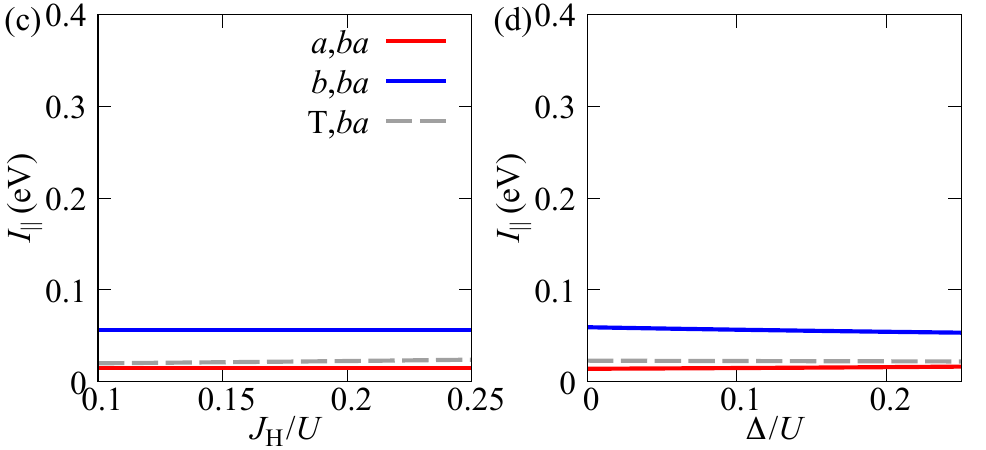}
\caption{$J_{\rm H}$ and  $\Delta$ dependencies of (a), (b) $J^{\xi\xi'}_{\parallel}$ and (c), (d) $I^{\xi,\bar{\nu}\nu}_{\parallel}$, where $t^{aa}_{\parallel}=0.117$~eV, $t^{bb}_{\parallel}=0.491$~eV, $t^{ab}_{\parallel}=0.242$~eV, $U=4$~eV, and $U' = U -2J_{\rm H}$.  
$\Delta/U=0.1$ is used in (a) and (c), while $J_{\rm H}/U=0.2$ is used in (b) and (d).}  
\label{fig7}
\end{center}
\end{figure}

\begin{figure}[t]
\begin{center}
\includegraphics[width=\columnwidth]{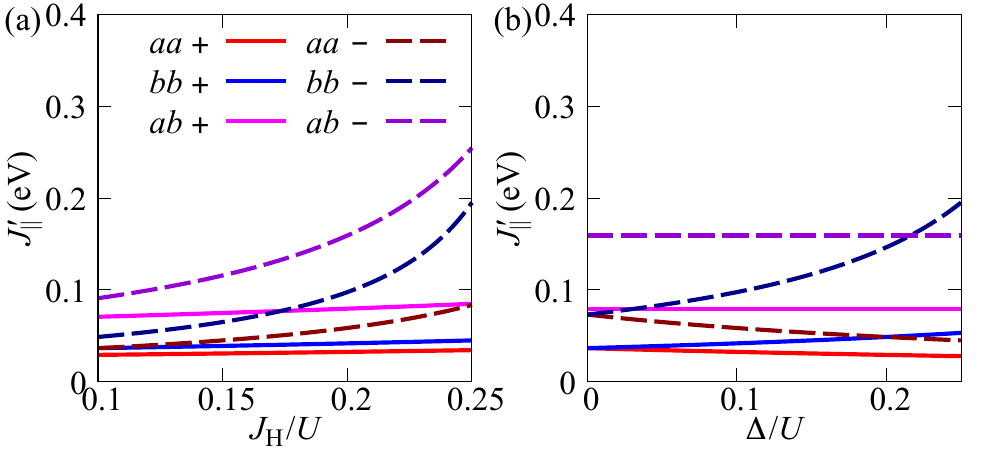} 
\includegraphics[width=\columnwidth]{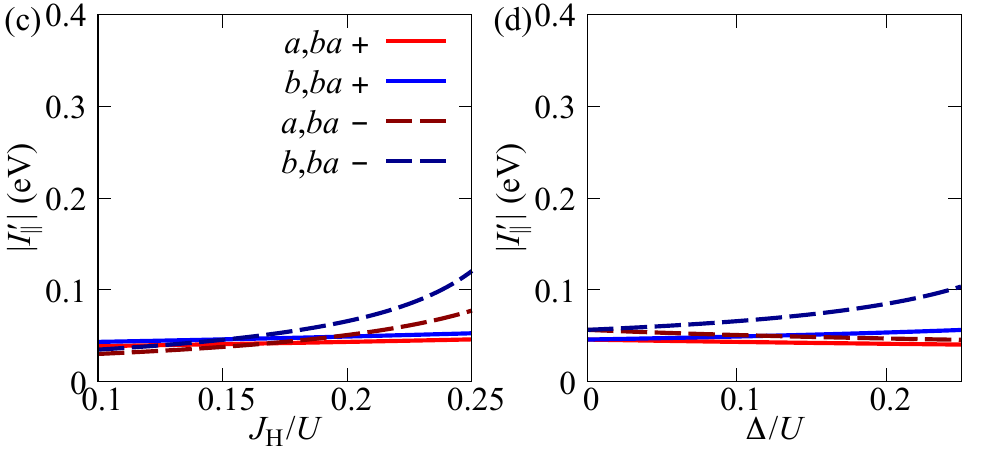}
\includegraphics[width=\columnwidth]{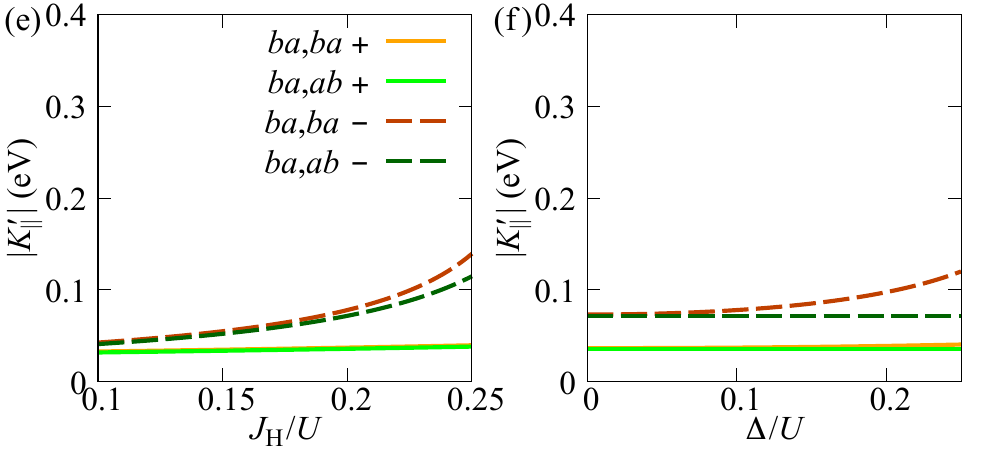}
\caption{$J_{\rm H}$ and  $\Delta$ dependencies of (a), (b) $J'^{\mu\nu}_{\parallel;\pm}$, (c), (d) $I'^{\mu,\bar{\nu}\nu}_{\parallel,\pm}$, and  (e), (f) $K'^{\bar{\mu}\mu,\bar{\nu}\nu}_{\parallel}$. 
$\Delta/U=0.1$ is used in (a), (c), and (e), while $J_{\rm H}/U=0.2$ is used in (b), (d), and (f). 
The other parameters are the same as in Fig.~\ref{fig7}.} 
\label{fig8}
\end{center}
\end{figure}

Since the interorbital hopping $t^{ab}_{\parallel}$ is nonzero, the values of the effective interactions that contain $t^{ab}_{\parallel}$ depend on the energy level difference $\Delta$ (see Appendix~\ref{appendix_E}). 
Figure~\ref{fig7} shows $J_{\rm H}$ and $\Delta$ dependencies of $J^{\xi\xi'}_{\parallel}$ and $I^{\xi,\bar{\nu}\nu}_{\parallel}$.  
Here, we employ $t^{aa}_{\parallel}= 0.117$~eV, $t^{bb}_{\parallel}= 0.491$~eV, and $t^{ab}_{\parallel}= 0.242$~eV evaluated in Ref.~\cite{HSakakibara2024_327}. 
Because the onsite energy $\Delta$ evaluated by first-principles calculations contains Coulomb contributions that are also included in the Hubbard model, we treat $\Delta$ as a parameter. 
Since $t^{bb}_{\parallel}$ is the largest, $J^{bb}_{\parallel}$, $J^{b{\rm T}}_{\parallel}$, and $J^{\rm TT}_{\parallel}$ become the leading spin interactions.  
On the other hand, $J^{aa}_{\parallel}$ is small due to small $t^{aa}_{\parallel}$. 
The correlated interorbital hoppings of  $I^{\xi,\bar{\nu}\nu}_{\parallel}$ are less than 0.1~eV, indicating that their contributions are small.   
Figure~\ref{fig8} shows $J'^{\mu\nu}_{\parallel}$, $I'^{\mu, \bar{\nu}\nu}_{\parallel}$, and $K'^{\bar{\mu}\mu,\bar{\nu}\nu}_{\parallel}$. 
Similarly to the interlayer interaction $J'^{ab}_{\perp;\pm}$, $J'^{\mu \nu}_{\parallel;-} > J'^{\mu \nu}_{\parallel;+}$ due to Hund's coupling $J_{\rm H}$ [see Fig.~\ref{fig8}(a)], indicating that the $J'^{\mu \nu}_{\parallel}$ terms in $\hat{\mathcal{H}}'^{(2)}_{{\rm eff}; \parallel}$ act as the ferromagnetic interaction between the spin-$\frac{1}{2}$ states. 
Because $t^{bb}_{\parallel}$ is the largest, the interorbital interaction $J'^{ab}_{\parallel;-}$ [$=(|t^{aa}_{\parallel}|^2+|t^{bb}_{\parallel}|^2)/(U' - J_{\rm H})$] is the largest in Fig.~\ref{fig8}(a).  
As seen in $J'^{bb}_{\parallel;\pm}$ [Fig.~\ref{fig8}(b)], $\Delta$ can lead to an effect similar to $J_{\rm H}$, where the energy difference between $J'^{bb}_{\parallel;+}$ and $J'^{bb}_{\parallel;-}$ increases as $\Delta$ increases.  
On the other hand, most values of $|I'^{\mu,\bar{\nu}\nu}_{\parallel}|$ and $|K'^{\bar{\mu}\mu,\bar{\nu}\nu}_{\parallel}|$ are less than 0.1~eV.

\subsection{Simplified $t$-$J$ model} \label{sec:model_LNO_C}

We propose a simplified effective Hamiltonian that incorporates the leading spin interactions in bilayer nickelate superconductors.   
The Hamiltonian of the simplified $t$-$J$ model is 
\begin{align}
\hat{\mathcal{H}} = \hat{\mathcal{H}}_0 + \hat{\mathcal{T}}'_0 + \hat{\mathcal{H}}_{J_{\perp}} + \hat{\mathcal{H}}_{J_{\parallel}}. 
\label{eq:simplified_tJ}
\end{align}
$\hat{\mathcal{H}}_0$ describes the energy of the onsite state, where  
\begin{align}
\hat{\mathcal{H}}_0 
= \Delta \sum_{j,l} \hat{\tilde{n}}_{j,l,b} 
+ \left( U' - J_{\rm H} + \Delta \right)\sum_{j,l} \hat{\tilde{n}}_{j,l,{\rm T}}  . 
\end{align}
$\hat{\mathcal{T}}'_0$ is the hopping term that does not change the number of doubly occupied orbitals and the number of empty sites [see Fig.~\ref{fig3}(a)].  $\hat{\mathcal{T}}'_0$ is given by 
\begin{align}
&\hat{\mathcal{T}}'_0 = 
-t^{aa}_{\perp}\sum_{j} \sum_{\sigma} 
\left(
\hat{\tilde{n}}_{j,1,{\rm T}} 
\hat{c}^{\dag}_{j,1,a,\sigma} \hat{c}_{j,2,a,\sigma}  
\hat{\tilde{n}}_{j,2,{\rm T}} 
+{\rm H.c.} \right)
\notag \\
&-\sum_{\mu,\nu} t^{\mu\nu}_{\parallel}
\sum_{\langle i, j \rangle} \sum_{l} \sum_{\sigma} 
\left(
\hat{\tilde{n}}_{i,l,{\rm T}} 
\hat{c}^{\dag}_{i,l,\mu,\sigma} \hat{c}_{j,l,\nu,\sigma}  
\hat{\tilde{n}}_{j,l,{\rm T}} 
+{\rm H.c.} \right), 
\end{align}
where both interorbital and intraorbital hoppings are included. 
$\hat{\mathcal{H}}_{J_{\perp}}$ represents the interlayer spin interaction.  
Incorporating the leading spin interactions attributed to $t^{aa}_{\perp}$, i.e., interlayer hopping between the $d_{3z^2-r^2}$ orbitals, $\hat{\mathcal{H}}_{J_{\perp}}$ is given by 
\begin{align}
\hat{\mathcal{H}}_{J_{\perp}}
&=J^{aa}_{\perp} \sum_{j} 
\left( 
\hat{\tilde{\bm{s}}}_{j,1,a} \cdot \hat{\tilde{\bm{s}}}_{j,2,a} 
- \frac{1}{4} \hat{\tilde{n}}_{j,1,a}\hat{\tilde{n}}_{j,2,a}
\right) 
\notag \\
&+J^{a {\rm T}}_{\perp} 
\sum_{j} 
\left( 
\hat{\tilde{\bm{s}}}_{j,1,a} \cdot \hat{\tilde{\bm{S}}}_{j,2} 
- \frac{1}{2} \hat{\tilde{n}}_{j,1,a}\hat{\tilde{n}}_{j,2,{\rm T}} 
\right) 
\notag \\
&+J^{a {\rm T}}_{\perp} 
\sum_{j} 
\left( 
\hat{\tilde{\bm{S}}}_{j,1} \cdot \hat{\tilde{\bm{s}}}_{j,2,a} 
- \frac{1}{2} \hat{\tilde{n}}_{j,1,{\rm T}}\hat{\tilde{n}}_{j,2,a}
\right) 
\notag \\
&+J^{{\rm TT}}_{\perp}
\sum_{j}
\left( 
\hat{\tilde{\bm{S}}}_{j,1} \cdot \hat{\tilde{\bm{S}}}_{j,2} 
- \hat{\tilde{n}}_{j,1,{\rm T}}\hat{\tilde{n}}_{j,2,{\rm T}} \right)
\notag \\
&-J'^{ab}_{\perp} \sum_{j} \sum_{\mu}
\left(
\hat{\tilde{\bm{s}}}_{j,1,\mu} \cdot \hat{\tilde{\bm{s}}}_{j,2,\bar{\mu}} 
+ \frac{\alpha_{\perp}}{4} \hat{\tilde{n}}_{j,1,\mu} \hat{\tilde{n}}_{j,2,\bar{\mu}} 
\right). 
\end{align}
The last term is given by integrating the $J'^{ab}_{\perp;+}$ and $ J'^{ab}_{\perp;-}$ terms in Eq.~(\ref{eq:Heff_perp_KK}), where $J'^{ab}_{\perp} = J'^{ab}_{\perp;-} - J'^{ab}_{\perp;+}$ and $\alpha_{\perp} = (3J'^{ab}_{\perp;-} + J'^{ab}_{\perp;+})/(J'^{ab}_{\perp;-} - J'^{ab}_{\perp;+})$. 
Because the antiferromagnetic interaction of $J'^{ab}_{\perp;+}$ and the ferromagnetic interaction of $J'^{ab}_{\perp;-}$ are competing in Eq.~(\ref{eq:Heff_perp_KK}), the $J'^{ab}_{\perp}$ term can be omitted from the simplified model if $J'^{ab}_{\perp;-} - J'^{ab}_{\perp;+}$ is small. 
Since $J'^{ab}_{\perp} = J'^{ab}_{\perp;-} - J'^{ab}_{\perp;+} > 0$ increases as $J_{\rm H}/U$ increases [see in Fig.~\ref{fig6}(b)], we should consider the $J'^{ab}_{\perp}$ term if Hund's coupling ($J_{\rm H}/U$) is sufficiently large. 
In addition, $\hat{\mathcal{H}}_{J_{\parallel}}$ incorporating the leading intralayer spin interactions is given by
\begin{align}
\hat{\mathcal{H}}_{J_{\parallel}}
&=J^{bb}_{\parallel} 
\sum_{\langle i, j \rangle} \sum_{l} 
\left( 
\hat{\tilde{\bm{s}}}_{i,l,b} \cdot \hat{\tilde{\bm{s}}}_{j,l,b} 
-\frac{1}{4} \hat{\tilde{n}}_{i,l,b}\hat{\tilde{n}}_{j,l,b}
\right) 
\notag \\
&+J^{b {\rm T}}_{\parallel} 
\sum_{\langle i, j \rangle} \sum_{l} 
\left( 
\hat{\tilde{\bm{s}}}_{i,l,b} \cdot \hat{\tilde{\bm{S}}}_{j,l} 
-\frac{1}{2} \hat{\tilde{n}}_{i,l,b}\hat{\tilde{n}}_{j,l,{\rm T}} 
\right) 
\notag \\
&+J^{b {\rm T}}_{\parallel} 
\sum_{\langle i, j \rangle} \sum_{l}
\left( 
\hat{\tilde{\bm{S}}}_{i,l} \cdot \hat{\tilde{\bm{s}}}_{j,l,b} 
-\frac{1}{2} \hat{\tilde{n}}_{i,l,{\rm T}}\hat{\tilde{n}}_{j,l,b}
\right) 
\notag \\
&+J^{{\rm TT}}_{\parallel}
\sum_{\langle i, j \rangle} \sum_{l}
\left( 
\hat{\tilde{\bm{S}}}_{i,l} \cdot \hat{\tilde{\bm{S}}}_{j,l} 
-\hat{\tilde{n}}_{i,l,{\rm T}}\hat{\tilde{n}}_{j,l,{\rm T}} 
\right)
\notag \\
&-J'^{ab}_{\parallel} \sum_{\langle i,j \rangle} \sum_{l} \sum_{\mu}
\left( 
\hat{\tilde{\bm{s}}}_{i,l,\mu} \cdot \hat{\tilde{\bm{s}}}_{j,l,\bar{\mu}} 
+\frac{\alpha_{\parallel}}{4} \hat{\tilde{n}}_{i,l,\mu}\hat{\tilde{n}}_{j,l,\bar{\mu}}
\right). 
\end{align}
Because $t^{bb}_{\parallel}$, i.e., intralayer hopping between the $d_{x^2-y^2}$ orbitals, has the largest magnitude, we take into account $J^{bb}_{\parallel}$, $J^{b{\rm T}}_{\parallel}$, $J^{\rm TT}_{\parallel}$, and $J'^{ab}_{\parallel}$ that contain $|t^{bb}_{\parallel}|^2$ (see Appendix~\ref{appendix_E}). 
As with the $J'^{ab}_{\perp}$ term, $J'^{ab}_{\parallel} = J'^{ab}_{\parallel;-} - J'^{ab}_{\parallel;+}$ and $\alpha_{\parallel} = (3J'^{ab}_{\parallel;-} + J'^{ab}_{\parallel;+})/(J'^{ab}_{\parallel;-} - J'^{ab}_{\parallel;+})$. 
We should consider the $J'^{ab}_{\parallel}$ term when $J'^{ab}_{\parallel;-} - J'^{ab}_{\parallel;+}$, i.e., $J_{\rm H}/U$, is sufficiently large.   
We do not include the correlated interorbital hopping terms in the simplified Hamiltonian because they are smaller than the leading spin interactions.

\subsection{Comparison with other $t$-$J$ models}

We comment on the relationship between our model and other $t$-$J$ models used for multilayer nickelate superconductors~\cite{HLange2024,XZQu2024,CLu2024_1,ZPan2024,JChen2024,JXZhang2024,HSchlomer2024,MBejas2025,YTian2025,PBorchia_arXiv,CLu_arXiv,MKakoi2024_DMRG,ZLuo2024,CLu2024_2,XZQu_arXiv,ZShao_arXiv,GDuan_arXiv,HOh2023,JRXue2024,HYang2024,HOh2025,HYang2025}. 
A simple model previously used is the single-orbital $t$-$J$ model only composed of the $d_{x^2-y^2}$ orbitals~\cite{HLange2024,XZQu2024,CLu2024_1,ZPan2024,JChen2024,JXZhang2024,HSchlomer2024,MBejas2025,YTian2025,PBorchia_arXiv,CLu_arXiv}. 
This single-orbital $t$-$J$ model is based on the assumption that the $d_{3z^2-r^2}$-orbital network is half-filling and electron mobility in the $d_{3z^2-r^2}$-orbital network is less significant. 
In the large-$J_{\rm H}$ limit with this assumption, the strong interlayer spin interaction between the $d_{3z^2-r^2}$ orbitals transfers to the interlayer spin interaction between the $d_{x^2-y^2}$ orbitals via the interorbital ferromagnetic interaction $J_{\rm H}$~\cite{CLu2024_1}. 
Since the interlayer electron hopping $t_{\perp}$ of the $d_{x^2-y^2}$ orbital is nearly zero, the single-orbital $t_{\parallel}$-$J_{\parallel}$-$J_{\perp}$ model~\cite{XZQu2024} is often used.  
However, the electron filling of the $d_{3z^2-r^2}$ orbital in La$_3$Ni$_2$O$_7$ is possibly less than half, and orbital hybridization effects due to the intralayer hopping $t^{ab}_{\parallel}$ are not negligible. 
To this end, the pairing properties and orbital hybridization effects with the $d_{3z^2-r^2}$ orbital have been investigated in the two-orbital $t$-$J$ models~\cite{MKakoi2024_DMRG,ZLuo2024,CLu2024_2,XZQu_arXiv,ZShao_arXiv,GDuan_arXiv}, where spins of the $d_{x^2-y^2}$ and $d_{3z^2-r^2}$ orbitals are defined independently regardless of the electron configurations of the two orbitals. 
Although these studies consider both $d_{x^2-y^2}$ and $d_{3z^2-r^2}$ orbitals, the two-orbital $t$-$J$ models in the previous studies do not use the spin interactions derived from the spin multiples in the two-orbital Hubbard model with Hund's coupling. 
In contrast to the previous two-orbital $t$-$J$ models~\cite{MKakoi2024_DMRG,ZLuo2024,CLu2024_2,XZQu_arXiv,ZShao_arXiv,GDuan_arXiv}, we introduce the effective model based on the spin multiplets, which are the exact single-site eigenstates in the presence of the spin-flip term in Hund's coupling. 

Our simplified $t$-$J$ model in Eq.~(\ref{eq:simplified_tJ}) is similar to the type-II $t$-$J$ model~\cite{HOh2023,JRXue2024,HYang2024}, which is defined by spin-$\frac{1}{2}$ (single spin) and spin-1 (spin triplet) in the large-$J_{\rm H}$ limit. 
As shown in Sec.~\ref{sec:model_LNO_C}, our effective model derived in Sec.~\ref{sec:model_I} can be reduced to the type-II $t$-$J$ model when only significant spin interactions are considered (where the correlated interorbital hopping terms are neglected). 
We comprehensively derive the two-site interactions from the two-orbital Hubbard model and present the effective Hamiltonian that includes the type-II $t$-$J$ model. 
In other words, we can selectively assemble the effective Hamiltonian. 
For example, if the correlated hopping terms ($I_{\parallel}$, $I'_{\parallel}$, and $K'_{\parallel}$) become essential, we can incorporate them into the effective model.   
Hence, our generalized $t$-$J$ model is useful for various strongly correlated two-orbital systems.


\section{Summary} \label{sec:summary}

We derived a $t$-$J$ model from the two-orbital Hubbard model and adapted it to the bilayer nickelate La$_3$Ni$_2$O$_7$.  
We listed the eigenenergies on a single site with two orbitals and classified the hopping processes using the projection operators to evaluate the energy changes associated with single-particle hopping. 
Then, using the Schrieffer-Wolff transformation, we derived the effective spin interactions when doubly occupied orbitals are excluded.   
We also showed that the Kugel-Khomskii--type interactions are derived by excluding empty sites. 
Furthermore, we assessed the effective spin interactions based on the hopping parameters evaluated in La$_3$Ni$_2$O$_7$, where we discussed the interlayer and intralayer spin interactions.  
Considering the obtained effective interactions, we proposed a simplified $t$-$J$ model for bilayer nickelate superconductors.  

One advantage of using the $t$-$J$ model is the reduced computational cost. 
While the 16 ($=4^2$) states must be taken into account on each site in the two-orbital Hubbard model, the number of local states can be reduced to seven ($\uparrow_a$, $\downarrow_a$, $\uparrow_b$, $\downarrow_b$, T$_{+}$, T$_0$, T$_-$) in the $t$-$J$ model (when $U$, $U'$, and $J_{\rm H}$ are sufficiently larger than the hopping parameters).  
Reducing the numerical cost enables us to access larger-size systems with high numerical accuracy using many-body solvers such as methods based on tensor network techniques. 
Numerical studies of the derived effective model are important for elucidating the mechanism of superconductivity and density wave states in multilayer nickelates.


\begin{acknowledgments}
We thank R. Ueda for fruitful discussions. 
This work was supported by Grants-in-Aid for Scientific Research from JSPS, KAKENHI Grants No.~JP20H01849, No.~JP24K06939, No.~JP24H00191 (T.K.), No.~JP22K04907 (K.K.), and No.~JP24K01333. 
M.K. was supported by JST SPRING (Grant No.~JPMJSP2138), the Kato Foundation for Promotion of Science (Grant No.~KS-3614), and JSPS Research Fellowships for Young Scientists (Grant No.~JP25KJ1758). 
\end{acknowledgments}


\appendix 


\section{Pair-hopping term} \label{appendix_A}

Here, we note the roles of the pair-hopping term 
\begin{align}
\hat{H}_{\rm P} = J_{\rm P} \sum_{\bm{R}} \left( \hat{c}^{\dag}_{\bm{R},a,\uparrow} \hat{c}^{\dag}_{\bm{R},a,\downarrow} \hat{c}_{\bm{R},b,\downarrow} \hat{c}_{\bm{R},b,\uparrow} +  {\rm H.c.} \right). 
\end{align}
$J_{\rm P}$ is the strength of the pair hopping, and we assume $J_{\rm P}>0$ in the following discussion. 
This pair-hopping term does not change the $\sigma_{\mu}$, S, and T states used in the ground-state configuration of the effective system that excludes doubly occupied orbitals. 
However, $J_{\rm P}$ partially modifies the energy spectrum of the excited states. 
Specifically, the eigenstates composed of the doubly occupied orbitals $\ket{{\rm D}_a} = \hat{c}^{\dag}_{a,\uparrow}\hat{c}^{\dag}_{a,\downarrow} \ket{0}$ and $\ket{{\rm D}_b} = \hat{c}^{\dag}_{b,\uparrow}\hat{c}^{\dag}_{b,\downarrow} \ket{0}$ are reorganized by the pair-hopping term because $\braket{{\rm D}_a|\hat{H}_{\rm P}|{\rm D}_b} = \braket{{\rm D}_b|\hat{H}_{\rm P}|{\rm D}_a} = J_{\rm P}$. 
When $J_{\rm P} \ne 0$, the matrix    
\begin{align}
\left( 
\begin{array}{cc}
U & J_{\rm P} \\
J_{\rm P} & U + 2\Delta
\end{array}
\right)
\end{align}
defined by $\ket{{\rm D}_a}$ and $\ket{{\rm D}_b}$ gives the reorganized eigenenergies 
\begin{align}
E_{\pm} = U + \Delta \pm \sqrt{\Delta^2 + J^2_{\rm P}} . 
\end{align}
Their eigenstates are given by 
\begin{align}
&\ket{{\rm D}_+} = u \ket{{\rm D}_b} + v \ket{{\rm D}_a},
\\
&\ket{{\rm D}_-} = - v \ket{{\rm D}_b} + u \ket{{\rm D}_a}
\end{align}
with 
\begin{align}
u = \sqrt{\frac{1}{2} \left( 1 \!+\! \frac{\Delta}{\sqrt{\Delta^2 \!+\! J^2_{\rm P}}} \right)}, \;\;
v = \sqrt{\frac{1}{2} \left( 1 \!-\! \frac{\Delta}{\sqrt{\Delta^2 \!+\! J^2_{\rm P}}} \right)}. 
\label{eq:PH_uv}
\end{align} 
Hence, $U$ for D$_{a}$ and $U+2\Delta$ for D$_b$ in Table~\ref{table1} are replaced by $U + \Delta \pm \sqrt{\Delta^2 + J^2_{\rm P}}$ for D$_{\pm}$. 
This indicates that the formulas of the coupling constants, e.g., $J^{\xi\xi'}_{\bm{R}\bm{R}'}$, in the effective model are partially modified by the pair hopping $J_{\rm P}$. 
If $J_{\rm P} \gg \Delta$, the eigenenergies and eigenstates can be simplified as $E_{\pm} \simeq U + \Delta \pm J_{\rm P}$ and $\ket{{\rm D}_{\pm}} \simeq (\ket{{\rm D}_a} \pm \ket{{\rm D}_b})/\sqrt{2}$. 

If we derive an effective model incorporating effects of $J_{\rm P}$, the single-particle hopping term must be further subdivided with the projection operators for $\ket{{\rm D}_+}$ and $\ket{{\rm D}_-}$.   
Using $u$ and $v$ in Eq.~(\ref{eq:PH_uv}), the projection operators for $\ket{{\rm D}_+}$ and $\ket{{\rm D}_-}$ are given by 
\begin{align}
\hat{P}_{\bm{R}}({\rm D}_+) 
&= u^2 \hat{P}_{\bm{R}}({{\rm D}_b}) + v^2 \hat{P}_{\bm{R}}({{\rm D}_a}) + uv \hat{Q}^{({\rm P})}_{\bm{R}}, 
\\
\hat{P}_{\bm{R}}({\rm D}_-) 
&= v^2 \hat{P}_{\bm{R}}({{\rm D}_b}) + u^2 \hat{P}_{\bm{R}}({{\rm D}_a}) - uv \hat{Q}^{({\rm P})}_{\bm{R}}, 
\end{align} 
respectively, where $\hat{Q}^{({\rm P})}_{\bm{R}} = \hat{c}^{\dag}_{\bm{R},a,\uparrow} \hat{c}^{\dag}_{\bm{R},a,\downarrow} \hat{c}_{\bm{R},b,\downarrow} \hat{c}_{\bm{R},b,\uparrow} + {\rm H.c.}$ is the pair-hopping operator.  
These D$_+$ and D$_-$ states appear after the hopping process $t^{\mu\nu}_{\bm{R}\bm{R}'} \hat{c}^{\dag}_{\bm{R},\mu,\sigma} \hat{c}_{\bm{R}',\nu,\sigma} \hat{P}_{\bm{R}}(\mu) \hat{P}_{\bm{R}'}(\xi')$ (where $\xi'=\nu, {\rm S}, {\rm T}$) occurs in $\hat{\mathcal{T}}_{+1}$ [Eq.~(\ref{eq:T+})]. 
To distinguish this process, it is necessary to decompose $\hat{\mathcal{T}}_{+1}$ into $\hat{\mathcal{T}}^{({\rm D})}_{+1}$ composed of $t^{\mu\nu}_{\bm{R}\bm{R}'}  \hat{c}^{\dag}_{\bm{R},\mu,\sigma} \hat{c}_{\bm{R}',\nu,\sigma} \hat{P}_{\bm{R}}(\mu) \hat{P}_{\bm{R}'}(\xi')$ and $\hat{\mathcal{T}}^{({\rm O})}_{+1}$ composed of $t^{\mu\nu}_{\bm{R}\bm{R}'} \hat{c}^{\dag}_{\bm{R},\mu,\sigma} \hat{c}_{\bm{R}',\nu,\sigma} [\hat{P}_{\bm{R}}({\rm S}) + \hat{P}_{\bm{R}}({\rm T})] \hat{P}_{\bm{R}'}(\xi')$. 
Once the operator $\hat{\mathcal{S}}^{(1)}$ in the Schrieffer-Wolff transformation for $\hat{\mathcal{T}}^{({\rm D})}_{+1} $ is obtained, the remaining procedures are essentially the same as the model derivation without the pair-hopping term. 
The analytical formulas for the effective interactions may become complicated due to $u$ and $v$. 
As an alternative to deriving analytical formulas, it is possible to combine a numerical approach~\cite{BKim2012,KRiedl2019} to determine the quantities of the interactions. 

\clearpage


\begin{widetext}

\section{Details of $\hat{\mathcal{S}}^{(1)}$ and $\hat{\mathcal{H}}^{(2)}_{\rm eff}$} \label{appendix_B}

The local Hamiltonian $\hat{\mathcal{H}}_0$ and the hopping operator $\hat{c}^{\dag}_{\bm{R},\mu,\sigma} \hat{c}_{\bm{R}',\nu,\sigma}$ give 
\begin{align}
\left[ \hat{\mathcal{H}}_0, \hat{c}^{\dag}_{\bm{R},\mu,\sigma} \hat{c}_{\bm{R}',\nu,\sigma} \right] 
&= \Delta_{\mu\nu} 
\hat{c}^{\dag}_{\bm{R},\mu,\sigma} \hat{c}_{\bm{R}',\nu,\sigma}
+ U \hat{c}^{\dag}_{\bm{R},\mu,\sigma} \hat{c}_{\bm{R}',\nu,\sigma} 
\left( \hat{n}_{\bm{R},\mu,\bar{\sigma}} - \hat{n}_{\bm{R}',\nu,\bar{\sigma}} \right)
+ \left( U' - \frac{J_{\rm H}}{2} \right) \hat{c}^{\dag}_{\bm{R},\mu,\sigma} \hat{c}_{\bm{R}',\nu,\sigma}
\left( \hat{n}_{\bm{R},\bar{\mu}} - \hat{n}_{\bm{R}',\bar{\nu}} \right)
\notag \\
&-J_{\rm H} \left( \hat{c}^{\dag}_{\bm{R},\mu,\bar{\sigma}} \hat{c}_{\bm{R}',\nu,\sigma} \hat{s}^{\sigma}_{\bm{R},\bar{\mu}} 
+ \sigma \hat{c}^{\dag}_{\bm{R},\mu,\sigma} \hat{c}_{\bm{R}',\nu,\sigma} \hat{s}^{z}_{\bm{R},\bar{\mu}} \right)
+ J_{\rm H} \left( \hat{c}^{\dag}_{\bm{R},\mu,\sigma} \hat{c}_{\bm{R}',\nu,\bar{\sigma}} \hat{s}^{\bar{\sigma}}_{\bm{R}',\bar{\nu}}
+ \sigma \hat{c}^{\dag}_{\bm{R},\mu,\sigma} \hat{c}_{\bm{R}',\nu,\sigma}   \hat{s}^{z}_{\bm{R}',\bar{\nu}} \right), 
\end{align}
where $\Delta_{\mu\nu} = \Delta ( \delta_{\mu,b} - \delta_{\nu,b} )$. 
We define $\sigma=\uparrow=+$ and $\sigma=\downarrow=-$, e.g., $\hat{s}^{\sigma}_{\bm{R},\mu} = \hat{s}^{+}_{\bm{R},\mu}$ for $\sigma=\uparrow$ and $\hat{s}^{\sigma}_{\bm{R},\mu} = \hat{s}^{-}_{\bm{R},\mu}$ for $\sigma=\downarrow$. 
The last two terms with $J_{\rm H}$ include the spin-flip operations, such as $\hat{c}^{\dag}_{\bm{R},\mu,\bar{\sigma}} \hat{c}_{\bm{R}',\nu,\sigma} \hat{s}^{\sigma}_{\bm{R},\bar{\mu}}$. 
However, we can lead to the hopping term $\hat{c}^{\dag}_{\bm{R},\mu,\sigma} \hat{c}_{\bm{R}',\nu,\sigma}$ by applying the projection operators, e.g., 
\begin{align}
-&J_{\rm H} \left( \hat{c}^{\dag}_{\bm{R},\mu,\bar{\sigma}} \hat{c}_{\bm{R}',\nu,\sigma} \hat{s}^{\sigma}_{\bm{R},\bar{\mu}} 
+ \sigma \hat{c}^{\dag}_{\bm{R},\mu,\sigma} \hat{c}_{\bm{R}',\nu,\sigma} \hat{s}^{z}_{\bm{R},\bar{\mu}} \right)
\hat{P}_{\bm{R}}({\rm S})
= - \frac{3}{2} J_{\rm H} \hat{c}^{\dag}_{\bm{R},\mu,\sigma} \hat{c}_{\bm{R}',\nu,\sigma} \hat{P}_{\bm{R}} ({\rm S}), 
\\
-&J_{\rm H}  \left( \hat{c}^{\dag}_{\bm{R},\mu,\bar{\sigma}} \hat{c}_{\bm{R}',\nu,\sigma} \hat{s}^{\sigma}_{\bm{R},\bar{\mu}}  
+ \sigma \hat{c}^{\dag}_{\bm{R},\mu,\sigma} \hat{c}_{\bm{R}',\nu,\sigma} \hat{s}^{z}_{\bm{R},\bar{\mu}} \right)
\hat{P}_{\bm{R}}({\rm T}_{\zeta})
= \frac{1}{2} J_{\rm H} \hat{c}^{\dag}_{\bm{R},\mu,\sigma} \hat{c}_{\bm{R}',\nu,\sigma} \hat{P}_{\bm{R}} ({\rm T}_{\zeta}), 
\end{align}
where $\hat{c}^{\dag}_{\bm{R},\mu,\uparrow} \hat{c}_{\bm{R}',\nu,\uparrow} \hat{P}_{\bm{R}} ({\rm T}_{+}) = \hat{c}^{\dag}_{\bm{R},\mu,\downarrow} \hat{c}_{\bm{R}',\nu,\downarrow} \hat{P}_{\bm{R}} ({\rm T}_{-})=0$. 
Since $\bigl[ \hat{\mathcal{H}}_0, \hat{P}_{\bm{R}}(\xi) \bigr]=0$, we find
\begin{align}
&\left[ \hat{\mathcal{H}}_0, \hat{c}^{\dag}_{\bm{R},\mu,\sigma} \hat{c}_{\bm{R}',\nu,\sigma} \hat{P}_{\bm{R}}(\xi) \hat{P}_{\bm{R}'}(\xi') \right] 
= \Delta E(\xi, \xi')\hat{c}^{\dag}_{\bm{R},\mu,\sigma} \hat{c}_{\bm{R}',\nu,\sigma} \hat{P}_{\bm{R}}(\xi) \hat{P}_{\bm{R}'}(\xi'). 
\end{align}
$\Delta E(\xi,\xi')$ corresponds to the energy change associated with the single-particle hopping and is summarized in Table~\ref{table2}. 
From this relation, we can prove that $i \hat{\mathcal{S}}^{(1)}_{+1}$ in Eq.~(\ref{eq:SWT_S1_+1}) satisfies $\bigl[ \hat{\mathcal{H}}_0 , i\hat{\mathcal{S}}^{(1)}_{+1} \bigr] = \hat{\mathcal{T}}_{+1}$.  

Once we find $i\hat{\mathcal{S}}^{(1)}_{\pm 1}$, we can derive the effective Hamiltonian in the second-order perturbation theory using Eq.~(\ref{eq:Heff_2nd}).  
When the model is defined on the space without doubly occupied orbitals, the operation $\hat{\mathcal{T}}_{-1} \hat{\mathcal{S}}^{(1)}_{+1}$ generates the two- and three-site interactions. 
Specifically, the two-site interactions are given by 
\begin{align}
\hat{\mathcal{H}}^{(2)}_{\rm eff;{\rm 2s}} 
=-\frac{1}{2}
\sum_{\bm{R},\bm{R}'} \sum_{\mu,\nu,\nu'} \sum_{\sigma,\sigma'} \sum_{\xi_1,\xi'_1,\xi_2,\xi'_2} 
\frac{t^{\nu'\mu}_{\bm{R}'\bm{R}}t^{\mu\nu}_{\bm{R}\bm{R}'}}{\Delta E(\xi_1, \xi'_1)} 
\hat{P}_{\bm{R}'}(\xi'_2)  \hat{P}_{\bm{R}}(\xi_2)  \hat{c}^{\dag}_{\bm{R}',\nu',\sigma'} \hat{c}_{\bm{R},\mu,\sigma'} 
\hat{c}^{\dag}_{\bm{R},\mu,\sigma} \hat{c}_{\bm{R}',\nu,\sigma} \hat{P}_{\bm{R}}(\xi_1)  \hat{P}_{\bm{R}'}(\xi'_1)
+ {\rm H.c.}, 
\label{eq:Heff_2nd_2site}
\end{align}
and three-site interactions are given by 
\begin{align}
\hat{\mathcal{H}}^{(2)}_{\rm eff;{\rm 3s}} 
= -\frac{1}{2}
\sum_{\bm{R},\bm{R}',\bm{R}''} \sum_{\mu,\nu,\nu''} \sum_{\sigma,\sigma'} \sum_{\xi_1,\xi'_1,\xi_2,\xi''_2} 
\frac{t^{\nu''\mu}_{\bm{R}''\bm{R}}t^{\mu\nu}_{\bm{R}\bm{R}'}}{\Delta E(\xi_1, \xi'_1)} 
\hat{P}_{\bm{R}''}(\xi''_2) \hat{P}_{\bm{R}}(\xi_2)  \hat{c}^{\dag}_{\bm{R}'',\nu'',\sigma'} \hat{c}_{\bm{R},\mu,\sigma'} 
\hat{c}^{\dag}_{\bm{R},\mu,\sigma} \hat{c}_{\bm{R}',\nu,\sigma} \hat{P}_{\bm{R}}(\xi_1)  \hat{P}_{\bm{R}'}(\xi'_1)
+ {\rm H.c.} 
\label{eq:Heff_2nd_3site}
\end{align}

We can derive effective spin interactions using $\hat{c}^{\dag}_{3,\sigma'} \hat{c}_{2,\sigma'}  \hat{c}^{\dag}_{2,\sigma} \hat{c}_{1,\sigma} =\hat{c}^{\dag}_{3,\sigma'} \left[ \left( 1-\hat{n}_{2}/2\right)\delta_{\sigma',\sigma} - \hat{\bm{s}}_{2} \cdot \bm{\sigma}_{\sigma'\sigma} \right] \hat{c}_{1,\sigma}$. 
When the indices 1 and 3 are the same, $\sum_{\sigma,\sigma'}\hat{c}^{\dag}_{1,\sigma'} \hat{c}_{2,\sigma'} \hat{c}^{\dag}_{2,\sigma} \hat{c}_{1,\sigma} = -2 \left[ \hat{\bm{s}}_{1} \cdot \hat{\bm{s}}_{2} - \left( \hat{n}_{1}/2 \right) \left( 1-\hat{n}_{2}/2\right) \right]$.


\section{Interactions with spin-singlet sites} \label{appendix_C}

This appendix presents the interactions that include the spin-singlet (S) state. 
To describe the Hamiltonian uniformly, we introduce the operator $\hat{\bm{\Sigma}} = \sum_{\lambda,\lambda'} \left(\bm{\Sigma}\right)_{\lambda\lambda'} \ket{\lambda}\bra{\lambda'}$ ($\lambda,\lambda'=$T$_+$, T$_0$, T$_-$, S) characterized by the matrices
\begin{gather}
\Sigma^{x} \!=\! \frac{1}{\sqrt{2}}
\left(
\begin{array}{@{}c|c@{}}
\sqrt{2}S^x &
\begin{matrix}
-1 \\
 0 \\
 1 
\end{matrix}\\
\hline
\begin{matrix}
-1 & 0 & 1
\end{matrix}
 & 0
\end{array}
\right), 
\;\;
\Sigma^{y} \!=\! \frac{1}{\sqrt{2}}
\left(
\begin{array}{@{}c|c@{}}
\sqrt{2}S^y &
\begin{matrix}
i \\
0 \\
i 
\end{matrix}\\
\hline
\begin{matrix}
-i & 0 & -i
\end{matrix}
 & 0
\end{array}
\right), 
\;\;
\Sigma^{z} \!=\! 
\left(
\begin{array}{@{}c|c@{}}
S^z &
\begin{matrix}
0 \\
1 \\
0 
\end{matrix}\\
\hline
\begin{matrix}
0 & 1 & 0
\end{matrix}
 & 0
\end{array}
\right). 
\end{gather}
These matrices are expressed in the basis set $\{\ket{{\rm T}_+}, \ket{{\rm T}_0}, \ket{{\rm T}_-}, \ket{{\rm S}}\}$. 
Using these matrices, the interactions among the spin-triplet and spin-singlet states can be described by $\hat{\bm{\Sigma}}_{\bm{R}} \cdot \hat{\bm{\Sigma}}_{\bm{R}'}-1$. 
Similarly, the interactions between the singlet or triplet state and the spin-$\frac{1}{2}$ state can be described by $\hat{\bm{\Sigma}}_{\bm{R}} \cdot \hat{\bm{s}}_{\bm{R}',\nu}-\frac{1}{2}$. 

The two-site interactions that do not change the initial and final configurations are given by   
\begin{align}
\hat{\mathcal{H}}^{(2)}_{J,{\rm S}}
&= \sum_{\langle \bm{R},\bm{R}' \rangle} \sum_{\nu} 
\left[
J^{{\rm S}\nu}_{\bm{R}\bm{R}'} 
\hat{P}_{\bm{R}}({\rm S}) \hat{P}_{\bm{R}'}(\nu) 
\left( \hat{\bm{\Sigma}}_{\bm{R}} \cdot \hat{\bm{s}}_{\bm{R}',\nu}  - \frac{1}{2} \right) 
\hat{P}_{\bm{R}}({\rm S}) \hat{P}_{\bm{R}'}(\nu) 
+\left( \bm{R} \leftrightarrow \bm{R}' \right) 
\right]
\notag \\
&+ \sum_{\langle \bm{R},\bm{R}' \rangle}
J^{{\rm SS}}_{\bm{R}\bm{R}'}
\hat{P}_{\bm{R}}({\rm S})\hat{P}_{\bm{R}'}({\rm S}) 
\left(\hat{\bm{\Sigma}}_{\bm{R}}\cdot\hat{\bm{\Sigma}}_{\bm{R}'} - 1\right)
\hat{P}_{\bm{R}}({\rm S})\hat{P}_{\bm{R}'}({\rm S}) 
\notag\\
&+\sum_{\langle \bm{R},\bm{R}' \rangle} 
\left[
J^{{\rm ST}}_{\bm{R}\bm{R}'}
\hat{P}_{\bm{R}}({\rm S})\hat{P}_{\bm{R}'}({\rm T}) 
\left(\hat{\bm{\Sigma}}_{\bm{R}}\cdot\hat{\bm{\Sigma}}_{\bm{R}'} - 1\right)
\hat{P}_{\bm{R}}({\rm S})\hat{P}_{\bm{R}'}({\rm T})
+\left( \bm{R} \leftrightarrow \bm{R}' \right) 
\right]
\end{align}
with 
\begin{align}
&J^{{\rm S}\nu}_{\bm{R}\bm{R}'} 
=\sum_{\mu} 
\left( 
\frac{|t^{\mu\nu}_{\bm{R}\bm{R}'}|^2}{ U \!+\! U' \!-\! 2J_{\rm H} \!+\! \Delta_{\mu\nu}} 
+\frac{|t^{\mu\nu}_{\bm{R}\bm{R}'}|^2}{U \!-\! U' \!-\! J_{\rm H} \!-\! \Delta_{\mu\nu} } 
\right),
\\
&J^{\rm SS}_{\bm{R}\bm{R}'}
=\frac{|t^{aa}_{\bm{R}\bm{R}'}|^2+|t^{bb}_{\bm{R}\bm{R}'}|^2}{U \!-\! 3J_{\rm H}} 
+\frac{|t^{ab}_{\bm{R}\bm{R}'}|^2 }{U \!-\! 3J_{\rm H} \!+\! \Delta} 
+\frac{|t^{ab}_{\bm{R}\bm{R}'}|^2}{U \!-\! 3J_{\rm H} \!-\! \Delta} , 
\\
&J^{\rm ST}_{\bm{R}\bm{R}'} = \frac{|t^{aa}_{\bm{R}\bm{R}'}|^2 + |t^{bb}_{\bm{R}\bm{R}'}|^2}{U \!-\! J_{\rm H}} 
+\frac{|t^{ab}_{\bm{R}\bm{R}'}|^2}{U \!-\! J_{\rm H} \!+\! \Delta} 
+\frac{|t^{ab}_{\bm{R}\bm{R}'}|^2}{ U \!-\! J_{\rm H} \!-\! \Delta} . 
\end{align} 
Note that since $\hat{P}_{\bm{R}}({\rm S}) \hat{\bm{\Sigma}}_{\bm{R}} \hat{P}_{\bm{R}}({\rm S})=0$, $\hat{P}_{\bm{R}}({\rm S}) \hat{P}_{\bm{R}'}(\nu) \left( \hat{\bm{\Sigma}}_{\bm{R}} \cdot \hat{\bm{s}}_{\bm{R}',\nu} - \frac{1}{2} \right) \hat{P}_{\bm{R}}({\rm S}) \hat{P}_{\bm{R}'}(\nu) = -\frac{1}{2}\hat{P}_{\bm{R}}({\rm S}) \hat{P}_{\bm{R}'}(\nu)$ and $\hat{P}_{\bm{R}}({\rm S})\hat{P}_{\bm{R}'}({\rm T}) \left(\hat{\bm{\Sigma}}_{\bm{R}}\cdot\hat{\bm{\Sigma}}_{\bm{R}'} - 1\right)\hat{P}_{\bm{R}}({\rm S})\hat{P}_{\bm{R}'}({\rm T}) = -\hat{P}_{\bm{R}}({\rm S})\hat{P}_{\bm{R}'}({\rm T})$. 
$J^{{\rm S}\nu}_{\bm{R}\bm{R}'}$, $J^{{\rm SS}}_{\bm{R}\bm{R}'}$, and $J^{{\rm ST}}_{\bm{R}\bm{R}'}$ give the spin-0 -- spin-$\frac{1}{2}$, spin-0 -- spin-0, and spin-0 -- spin-1 interactions, respectively. 
The correlated interorbital hopping term corresponding to $I^{{\rm T},\bar{\nu}\nu}_{\bm{R}\bm{R}'}$ [Eq.~(\ref{eq:H_I_sT})] is given by 
\begin{align}
\hat{\mathcal{H}}^{(2)}_{I,{\rm S}}
&= \sum_{\langle \bm{R},\bm{R}' \rangle} \sum_{\nu} 
I^{{\rm S},\bar{\nu}\nu}_{\bm{R}\bm{R}'}\hat{P}_{\bm{R}}({\rm S}) \hat{P}_{\bm{R}'}(\bar{\nu})
\left[
\sum_{\sigma,\sigma'} \hat{c}^{\dag}_{\bm{R}',\bar{\nu},\sigma'}    
\left( \hat{\bm{\Sigma}}_{\bm{R}} \! \cdot \!  \frac{\bm{\sigma}_{\sigma'\sigma}}{2} \! - \! \frac{1}{2} \delta_{\sigma',\sigma}\right) 
\hat{c}_{\bm{R}',\nu,\sigma} 
\right]
\hat{P}_{\bm{R}}({\rm S}) \hat{P}_{\bm{R}'}(\nu)
+ \left( \bm{R} \leftrightarrow \bm{R}' \right) 
\end{align}
with
\begin{align}
I^{{\rm S},\bar{\nu}\nu}_{\bm{R}\bm{R}'} &=
\frac{1}{2} \sum_{\mu} \left(
\frac{t^{\bar{\nu}\mu}_{\bm{R}'\bm{R}}t^{\mu\nu}_{\bm{R}\bm{R}'}}{U \!+\! U' \!-\! 2J_{\rm H} \!+\! \Delta_{\mu\nu}}
+\frac{t^{\bar{\nu}\mu}_{\bm{R}'\bm{R}}t^{\mu\nu}_{\bm{R}\bm{R}'}}{U \!+\! U' \!-\! 2J_{\rm H} \!+\! \Delta_{\mu\bar{\nu}}} 
\right). 
\end{align}

In addition, the transitions between the spin-singlet and spin-triplet states (${\rm S} \leftrightarrow {\rm T}$) are possible, and the interactions including these transitions are given by 
\begin{align}
\hat{\mathcal{H}}^{(2)}_{X, {\rm S}}
&= \sum_{\langle \bm{R},\bm{R}' \rangle} \sum_{\nu}
\left[ X^{{\rm T}{\rm S},\nu\nu}_{\bm{R}\bm{R}'} 
\hat{P}_{\bm{R}}({\rm T}) \hat{P}_{\bm{R}'}(\nu)
\left( \bm{\Sigma}_{\bm{R}} \cdot \hat{\bm{s}}_{\bm{R}',\nu}
- \frac{1}{2}\right) 
\hat{P}_{\bm{R}}({\rm S}) \hat{P}_{\bm{R}'}(\nu)
+ \left( \bm{R} \leftrightarrow \bm{R}' \right)
+ {\rm H.c.} \right]
\notag \\
&+ \sum_{\langle \bm{R},\bm{R}' \rangle} \sum_{\nu}
\left\{ X^{{\rm T}{\rm S},\bar{\nu}\nu}_{\bm{R}\bm{R}'} 
\hat{P}_{\bm{R}}({\rm T}) \hat{P}_{\bm{R}'}(\bar{\nu})
\left[
\sum_{\sigma,\sigma'} \hat{c}^{\dag}_{\bm{R}',\bar{\nu},\sigma'}    
\left( \hat{\bm{\Sigma}}_{\bm{R}} \! \cdot \! \frac{\bm{\sigma}_{\sigma'\sigma}}{2} \! - \! \frac{1}{2} \delta_{\sigma',\sigma}\right) 
\hat{c}_{\bm{R}',\nu,\sigma} 
\right]
\hat{P}_{\bm{R}}({\rm S}) \hat{P}_{\bm{R}'}(\nu)
+ \left( \bm{R} \leftrightarrow \bm{R}' \right)
+ {\rm H.c.} \right\}
\notag \\
&+\sum_{\langle \bm{R},\bm{R}' \rangle}
\left[ 
X^{{\rm TS},{\rm TS}}_{\bm{R}\bm{R}'} 
\hat{P}_{\bm{R}}({\rm T})\hat{P}_{\bm{R}'}({\rm T}) 
\left(\hat{\bm{\Sigma}}_{\bm{R}}\cdot\hat{\bm{\Sigma}}_{\bm{R}'} - 1\right)
\hat{P}_{\bm{R}}({\rm S})\hat{P}_{\bm{R}'}({\rm S}) 
+ {\rm H.c.} \right]
\notag \\
&+\sum_{\langle \bm{R},\bm{R}' \rangle}
\left[ 
X^{{\rm TS},{\rm ST}}_{\bm{R}\bm{R}'} 
\hat{P}_{\bm{R}}({\rm T}) \hat{P}_{\bm{R}'}({\rm S})
\left(\hat{\bm{\Sigma}}_{\bm{R}}\cdot\hat{\bm{\Sigma}}_{\bm{R}'} - 1\right)
\hat{P}_{\bm{R}}({\rm S}) \hat{P}_{\bm{R}'}({\rm T})
+ \left( \bm{R} \leftrightarrow \bm{R}' \right)
\right] \notag \\
&+\sum_{\langle \bm{R},\bm{R}' \rangle}
\left[ 
X^{{\rm TS},{\rm TT}}_{\bm{R}\bm{R}'} 
\hat{P}_{\bm{R}}({\rm T})\hat{P}_{\bm{R}'}({\rm T}) 
\left(\hat{\bm{\Sigma}}_{\bm{R}}\cdot\hat{\bm{\Sigma}}_{\bm{R}'} - 1\right)
\hat{P}_{\bm{R}}({\rm S})\hat{P}_{\bm{R}'}({\rm T}) 
+ \left( \bm{R} \leftrightarrow \bm{R}' \right) + {\rm H.c.}
\right],
\end{align}
where 
\begin{align}
X^{{\rm T}{\rm S}, \nu \nu}_{\bm{R}\bm{R}'} &= 
\frac{1}{2} \sum_{\mu} (\delta_{\mu,a} - \delta_{\mu,b}) \left(
\frac{|t^{\mu\nu}_{\bm{R}\bm{R}'}|^2}{U \!+\! U' \!+\! \Delta_{\mu\nu}} 
+\frac{|t^{\mu\nu}_{\bm{R}\bm{R}'}|^2}{U \!-\! U' \!+\! J_{\rm H} \!+\! \Delta_{\nu\mu}} 
+\frac{|t^{\mu\nu}_{\bm{R}\bm{R}'}|^2}{U \!+\! U' \!-\! 2J_{\rm H} \!+\! \Delta_{\mu\nu}} 
+\frac{|t^{\mu\nu}_{\bm{R}\bm{R}'}|^2}{U \!-\! U' \!-\! J_{\rm H} \!+\! \Delta_{\nu\mu}} 
\right),\\
X^{{\rm T}{\rm S},\bar{\nu}\nu}_{\bm{R}\bm{R}'} &= 
\frac{1}{2} \sum_{\mu} (\delta_{\mu,a} - \delta_{\mu,b}) \left( 
\frac{t^{\bar{\nu}\mu}_{\bm{R}'\bm{R}}t^{\mu\nu}_{\bm{R}\bm{R}'}}{U \!+\! U' \!-\! 2J_{\rm H} \!+\! \Delta_{\mu\nu}} 
+\frac{t^{\bar{\nu}\mu}_{\bm{R}'\bm{R}}t^{\mu\nu}_{\bm{R}\bm{R}'}}{U \!+\! U' \!+\! \Delta_{\mu\bar{\nu}}}
\right),\\
X^{{\rm TS},{\rm TS}}_{\bm{R}\bm{R}'} &= 
\frac{1}{2} \sum_{\mu} \left[
\left(
\frac{|t^{\mu\mu}_{\bm{R}\bm{R}'}|^2}{U \!+\! J_{\rm H}}
-\frac{|t^{\mu\bar{\mu}}_{\bm{R}\bm{R}'}|^2}{U \!+\! J_{\rm H} \!+\! \Delta_{\mu\bar{\mu}}}
\right)
+ \left( 
\frac{|t^{\mu\mu}_{\bm{R}\bm{R}'}|^2}{U \!-\! 3J_{\rm H}}
-\frac{|t^{\mu\bar{\mu}}_{\bm{R}\bm{R}'}|^2}{U \!-\! 3J_{\rm H} \!+\! \Delta_{\mu\bar{\mu}}}
\right) 
\right],\\
X^{{\rm TS},{\rm ST}}_{\bm{R}\bm{R}'} &= 
\frac{|t^{aa}_{\bm{R}\bm{R}'}|^2+|t^{bb}_{\bm{R}\bm{R}'}|^2}{U \!-\! J_{\rm H}}
-\frac{|t^{ab}_{\bm{R}\bm{R}'}|^2}{U \!-\! J_{\rm H} \!+\! \Delta}
-\frac{|t^{ab}_{\bm{R}\bm{R}'}|^2}{U \!-\! J_{\rm H} \!-\! \Delta},
\\
X^{{\rm TS},{\rm TT}}_{\bm{R}\bm{R}'} &= 
\frac{1}{2} \left(
\frac{|t^{aa}_{\bm{R}\bm{R}'}|^2}{U \!+\! J_{\rm H}}
-\frac{|t^{bb}_{\bm{R}\bm{R}'}|^2}{U \!+\! J_{\rm H}}
\right)
+\frac{1}{2} \left(
\frac{|t^{aa}_{\bm{R}\bm{R}'}|^2}{U \!-\! J_{\rm H}}
-\frac{|t^{bb}_{\bm{R}\bm{R}'}|^2}{U \!-\! J_{\rm H}}
\right). 
\end{align}


\section{Details of $\hat{\mathcal{S}}'^{(1)}$ and $\hat{\mathcal{H}}'^{(2)}_{\rm eff}$} \label{appendix_D}

The hopping term $\hat{\mathcal{T}}'_{+1}$ in Eq.~(\ref{eq:T'+1}) is characterized by $\hat{P}_{\bm{R}}(\xi) \hat{c}^{\dag}_{\bm{R},\mu,\sigma} \hat{c}_{\bm{R}',\nu,\sigma}  \hat{P}_{\bm{R}'}(\nu)$ with $\xi = {\rm S}, {\rm T}$. 
For this hopping term, 
\begin{align}
&\left[ \hat{\mathcal{H}}_0, \hat{P}_{\bm{R}}(\xi) \hat{c}^{\dag}_{\bm{R},\mu,\sigma} \hat{c}_{\bm{R}',\nu,\sigma} \hat{P}_{\bm{R}'}(\nu) \right] 
= \Delta E'(\xi) \hat{P}_{\bm{R}}(\xi) \hat{c}^{\dag}_{\bm{R},\mu,\sigma} \hat{c}_{\bm{R}',\nu,\sigma}  \hat{P}_{\bm{R}'}(\nu), 
\label{eq:H0_hop}
\end{align}
where $\Delta E'(\xi={\rm S}) = U' + J_{\rm H} + \Delta_{\mu\nu}$ and $\Delta E'(\xi={\rm T}) = U' - J_{\rm H} + \Delta_{\mu\nu}$. 
Equation~(\ref{eq:H0_hop}) gives $i \hat{\mathcal{S}}'^{(1)}_{+1}$ in Eq.~(\ref{eq:SWT_S'1_+1}) satisfying $\bigl[ \hat{\mathcal{H}}_0 , i\hat{\mathcal{S}}'^{(1)}_{+1}  \bigr] = \hat{\mathcal{T}}'_{+1}$.  

In the second-order process, the first hopping creates an empty site at $\bm{R}'$, and the second hopping must annihilate the empty site at $\bm{R}'$. 
In the ground-state configuration without empty sites, three-site hoppings are not allowed in $\hat{\mathcal{H}}'^{(2)}_{\rm eff}$.  
Then, the two-site interactions are generated by 
\begin{align}
&\hat{\mathcal{H}}'^{(2)}_{\rm eff} 
= -\frac{1}{2}\sum_{\bm{R},\bm{R}'} \sum_{\mu, \mu', \nu, \nu'} \sum_{\sigma,\sigma'} \sum_{\xi}
\frac{t^{\nu'\mu'}_{\bm{R}'\bm{R}}t^{\mu\nu}_{\bm{R}\bm{R}'}}{\Delta E'(\xi)}
\left[
\hat{P}_{\bm{R}'}(\nu') 
\hat{c}^{\dag}_{\bm{R}',\nu',\sigma'} 
\hat{c}_{\bm{R},\mu',\sigma'}
\hat{P}_{\bm{R}}(\xi)
\right]
\left[
\hat{P}_{\bm{R}}(\xi) 
\hat{c}^{\dag}_{\bm{R},\mu,\sigma} \hat{c}_{\bm{R}',\nu,\sigma}
\hat{P}_{\bm{R}'}(\nu) 
\right]
+{\rm H.c.}
\end{align}
We can derive the effective Hamiltonians in Sec.~\ref{sec:model_II_C} using 
\begin{align}
&\sum_{\sigma,\sigma'} 
\left[
\hat{P}_{\bm{R}'}(\nu') 
\hat{c}^{\dag}_{\bm{R}',\nu',\sigma'} 
\hat{c}_{\bm{R},\mu',\sigma'}
\hat{P}_{\bm{R}}({\rm S})
\right]
\left[
\hat{P}_{\bm{R}}({\rm S}) 
\hat{c}^{\dag}_{\bm{R},\mu,\sigma} \hat{c}_{\bm{R}',\nu,\sigma}
\hat{P}_{\bm{R}'}(\nu) 
\right]
\\
&= \sum_{\sigma_1,\sigma_2, \sigma_3,\sigma_4} 
\! \left( \frac{1}{4} \delta_{\sigma_1,\sigma_2} \delta_{\sigma_3,\sigma_4} 
\!-\! \frac{1}{4} \bm{\sigma}_{\sigma_1\sigma_2} \cdot \bm{\sigma}_{\sigma_3\sigma_4} 
\right)
\hat{P}_{\bm{R}}(\bar{\mu}') \hat{P}_{\bm{R}'}(\nu') 
\hat{c}^{\dag}_{\bm{R},\bar{\mu}',\sigma_1} \hat{c}_{\bm{R},\bar{\mu},\sigma_2}
\hat{c}^{\dag}_{\bm{R}',\nu',\sigma_3} \hat{c}_{\bm{R}',\nu,\sigma_4} 
\hat{P}_{\bm{R}}(\bar{\mu}) \hat{P}_{\bm{R}'}(\nu) , 
\notag \\
& \sum_{\sigma,\sigma'} 
\left[
\hat{P}_{\bm{R}'}(\nu') 
\hat{c}^{\dag}_{\bm{R}',\nu',\sigma'} 
\hat{c}_{\bm{R},\mu',\sigma'}
\hat{P}_{\bm{R}}({\rm T})
\right]
\left[
\hat{P}_{\bm{R}}({\rm T}) 
\hat{c}^{\dag}_{\bm{R},\mu,\sigma} \hat{c}_{\bm{R}',\nu,\sigma}
\hat{P}_{\bm{R}'}(\nu) 
\right]
\\
&= \left( \delta_{\mu',\mu} \!-\! \delta_{\mu',\bar{\mu}} \right) \! 
\sum_{\sigma_1,\sigma_2,\sigma_3,\sigma_4} 
\! \left( \frac{3}{4} \delta_{\sigma_1,\sigma_2} \delta_{\sigma_3,\sigma_4} 
\!+\! \frac{1}{4} \bm{\sigma}_{\sigma_1\sigma_2} \cdot \bm{\sigma}_{\sigma_3\sigma_4} 
\right) 
\hat{P}_{\bm{R}}(\bar{\mu}') \hat{P}_{\bm{R}'}(\nu') 
\hat{c}^{\dag}_{\bm{R},\bar{\mu}',\sigma_1} \hat{c}_{\bm{R},\bar{\mu},\sigma_2}
\hat{c}^{\dag}_{\bm{R}',\nu',\sigma_3} \hat{c}_{\bm{R}',\nu,\sigma_4} 
\hat{P}_{\bm{R}}(\bar{\mu}) \hat{P}_{\bm{R}'}(\nu) . 
\notag 
\end{align}

\end{widetext}


\section{Effective interactions for bilayer nickelate superconductors} \label{appendix_E}

When $t^{aa}_{\perp}\ne 0$ and $t^{ab}_{\perp}=t^{bb}_{\perp}=0$, the interlayer spin interactions in $\hat{\mathcal{H}}^{(2)}_{{\rm eff}; \perp}$ [Eq.~(\ref{eq:Heff_perp})] and $\hat{\mathcal{H}}'^{(2)}_{{\rm eff}; \perp}$ [Eq.~(\ref{eq:Heff_perp_KK})] are given by 

\begin{align}
&J^{aa}_{\perp} = \frac{4|t^{aa}_{\perp}|^2}{U} , 
\\
&J^{{\rm T} a}_{\perp} = J^{a {\rm T}}_{\perp} 
=\frac{|t^{aa}_{\perp}|^2}{U \!+\! U'}
+\frac{|t^{aa}_{\perp}|^2}{U \!-\! U' \!+\! J_{\rm H}} , 
\\
&J^{\rm TT}_{\perp}
= \frac{|t^{aa}_{\perp}|^2}{U \!+\! J_{\rm H}} , 
\end{align}
and 
\begin{align}
&J'^{ab}_{\perp;\pm} = J'^{ba}_{\perp;\pm}
=\frac{|t^{aa}_{\perp}|^2}{U' \! \pm \!  J_{\rm H} }.  
\end{align}

In contrast to the interlayer hopping $t^{\mu\nu}_{\perp}$, all components of the intralayer hoppings $t^{aa}_{\parallel}$, $t^{bb}_{\parallel}$, and $t^{ab}_{\parallel}$ are considered in the two-orbital model for the bilayer nickelate La$_3$Ni$_2$O$_7$~\cite{HSakakibara2024_327}. 
The effective intralayer interactions in $\hat{\mathcal{H}}^{(2)}_{{\rm eff}; \parallel}$ [Eq.~(\ref{eq:Heff_para})] are given by 
\begin{align}
&J^{aa}_{\parallel} = 
\frac{4|t^{aa}_{\parallel}|^2}{U}, 
\;\;\; 
J^{bb}_{\parallel} = 
\frac{4|t^{bb}_{\parallel}|^2}{U},
\\
&J^{ab}_{\parallel} = J^{ba}_{\parallel} 
=\frac{2|t^{ab}_{\parallel}|^2}{U \!+\! \Delta} 
+\frac{2|t^{ab}_{\parallel}|^2}{U \!-\! \Delta}, 
\end{align}
\begin{align}
J^{{\rm T}a}_{\parallel} = J^{a {\rm T}}_{\parallel} 
&=\frac{|t^{aa}_{\parallel}|^2}{U \!+\! U'}  
+\frac{|t^{aa}_{\parallel}|^2}{U \!-\! U' \!+\! J_{\rm H}} 
\notag \\
&+\frac{|t^{ab}_{\parallel}|^2}{U \!+\! U' \!+\! \Delta} 
+\frac{|t^{ab}_{\parallel}|^2}{U \!-\! U' \!+\! J_{\rm H} \!-\! \Delta }, 
\\
J^{{\rm T}b}_{\parallel} = J^{b {\rm T}}_{\parallel} 
&=\frac{|t^{bb}_{\parallel}|^2}{U \!+\! U'}  
+\frac{|t^{bb}_{\parallel}|^2}{U \!-\! U' \!+\! J_{\rm H}} 
\notag \\
&+\frac{|t^{ab}_{\parallel}|^2}{U \!+\! U' \!-\! \Delta} 
+\frac{|t^{ab}_{\parallel}|^2}{U \!-\! U' \!+\! J_{\rm H} \!+\! \Delta}, 
\end{align}
\begin{align}
&J^{\rm TT}_{\parallel}
=\frac{|t^{aa}_{\parallel}|^2 + |t^{bb}_{\parallel}|^2}{U \!+\! J_{\rm H}} 
+\frac{|t^{ab}_{\parallel}|^2}{U \!+\! J_{\rm H} \!+\! \Delta} 
+\frac{|t^{ab}_{\parallel}|^2}{U \!+\! J_{\rm H} \!-\! \Delta},  
\end{align} 
\begin{align}
&I^{a,ba}_{\parallel} = I^{a,ab}_{\parallel} = I^{ba,a}_{\parallel} = I^{ab,a}_{\parallel} 
=\frac{t^{ab}_{\parallel} t^{aa}_{\parallel}}{U} 
+\frac{t^{ab}_{\parallel} t^{aa}_{\parallel}}{U \!-\! \Delta}  , 
\\
&I^{b,ba}_{\parallel} = I^{b,ab}_{\parallel} = I^{ba,b}_{\parallel} = I^{ab,b}_{\parallel} 
=\frac{t^{ab}_{\parallel} t^{bb}_{\parallel}}{U} 
+\frac{t^{ab}_{\parallel} t^{bb}_{\parallel}}{U \!+\! \Delta} , 
\end{align}
and 
\begin{align}
&I^{{\rm T},ba}_{\parallel} = I^{{\rm T},ab}_{\parallel} = I^{ba, {\rm T}}_{\parallel} = I^{ab,{\rm T}}_{\parallel} 
\notag \\
&=\frac{1}{2} 
\left[ 
\frac{t^{ab}_{\parallel} (t^{aa}_{\parallel} \!+\! t^{bb}_{\parallel})}{U \!+\! U'} 
+\frac{t^{ab}_{\parallel}t^{a a}_{\parallel}}{U \!+\! U' \!-\! \Delta} 
+\frac{t^{ab}_{\parallel}t^{bb}_{\parallel}}{U \!+\! U' \!+\! \Delta } 
\right]. 
\end{align}
The effective intralayer interactions in $\hat{\mathcal{H}}'^{(2)}_{{\rm eff}; \parallel}$ [Eq.~(\ref{eq:Heff_para_KK})] are given by 
\begin{align}
&J'^{aa}_{\parallel;\pm} 
=\frac{2|t^{a b}_{\parallel}|^2}{U' \pm J_{\rm H} \!+\! \Delta}, 
\;\;\; 
J'^{bb}_{\parallel;\pm} 
=\frac{2|t^{a b}_{\parallel}|^2}{U' \pm J_{\rm H} \!-\! \Delta}, 
\\
&J'^{ab}_{\parallel;\pm} = J'^{ba}_{\parallel;\pm} 
=\frac{|t^{aa}_{\parallel}|^2+|t^{bb}_{\parallel}|^2}{U' \! \pm \! J_{\rm H}}, 
\label{eq_J'ab_intra}
\end{align}
\begin{align}
&I'^{a,ba}_{\parallel;\pm} = I'^{a,ab}_{\parallel;\pm} = I'^{ba,a}_{\parallel;\pm} = I'^{ab,a}_{\parallel;\pm} 
\notag \\
&= \frac{1}{2} 
\Biggl[ 
\frac{t^{ab}_{\parallel} (t^{bb}_{\parallel} \pm t^{aa}_{\parallel})}{U' \! \pm \! J_{\rm H}}
+\frac{t^{ab}_{\parallel} (t^{bb}_{\parallel} \pm t^{aa}_{\parallel})}{U' \! \pm \! J_{\rm H} \!+\! \Delta}
\Biggr], 
\\
&I'^{b,ba}_{\parallel;\pm} = I'^{b,ab}_{\parallel;\pm} = I'^{ba,b}_{\parallel;\pm} = I'^{ab,b}_{\parallel;\pm} 
\notag \\
&=\frac{1}{2} 
\Biggl[
\frac{t^{ab}_{\parallel} (t^{aa}_{\parallel} \pm t^{bb}_{\parallel})}{U' \! \pm \! J_{\rm H}}
+\frac{t^{ab}_{\parallel} (t^{aa}_{\parallel} \pm t^{bb}_{\parallel})}{U' \! \pm \! J_{\rm H} \!-\! \Delta}
\Biggr], 
\end{align}
and
\begin{align}
&K'^{ba,ba}_{\parallel;\pm} = K'^{ab,ab}_{\parallel;\pm}
=\pm \frac{|t^{ab}_{\parallel}|^2}{U' \! \pm \! J_{\rm H} \!+\!  \Delta}
\pm \frac{|t^{ab}_{\parallel}|^2}{U' \! \pm \! J_{\rm H} \!-\! \Delta}, 
\\
&K'^{ba,ab}_{\parallel;\pm} = K'^{ab,ba}_{\parallel;\pm}
=\pm \frac{2t^{aa}_{\parallel} t^{bb}_{\parallel}}{U' \! \pm \! J_{\rm H}}. 
\end{align}


\bibliography{reference}

\end{document}